\documentclass[journal]{IEEEtran}

\ifCLASSINFOpdf
  \usepackage[pdftex]{graphicx}
  \graphicspath{{../} {../pdf/} {../jpeg/}}
  \DeclareGraphicsExtensions{.pdf,.jpeg}
\else
  \usepackage[dvips]{graphicx}
  \graphicspath{{../} {../eps/}}
  \DeclareGraphicsExtensions{.eps}
\fi
\usepackage{amsmath}
\usepackage{amssymb}
\usepackage[linesnumbered,ruled,vlined]{algorithm2e}
\usepackage{array}
\usepackage{multirow}
\usepackage{booktabs}
\usepackage{lipsum}
\usepackage{bm}
\usepackage{color}
\usepackage{booktabs}
\usepackage{footnote}
\usepackage{tabularx}
\usepackage{float}
\usepackage[flushleft]{threeparttable}
\usepackage[dvipsnames,table]{xcolor}
\usepackage[normalem]{ulem}
\usepackage{url}

\usepackage{nomencl}
\makenomenclature
\usepackage{etoolbox}
\renewcommand\nomgroup[1]{%
  \item[\bfseries
  \ifstrequal{#1}{A}{\textit{Abbreviations}}{%
  \ifstrequal{#1}{U}{\textit{Variables for Uncertainties}}{%
  \ifstrequal{#1}{F}{\textit{Variables and Parameters for the PPF Formulation}}{%
  \ifstrequal{#1}{L}{\textit{Variables and Parameters for the LRA Method}}{}}}}%
]}

\begin{document}

\bstctlcite{IEEEbib:BSTcontrol}

\title{Probabilistic Power Flow Calculation using Non-intrusive Low-rank Approximation Method}

\author{Hao~Sheng,~\IEEEmembership{Member,~IEEE,}
        and~Xiaozhe~Wang,~\IEEEmembership{Member,~IEEE,}
\thanks{This work is supported by Natural Sciences and Engineering Research Council (NSERC) under Discovery Grant NSERC RGPIN-2016-04570.}
\thanks{The authors are with the Department of Electrical and Computer Engineering, McGill University, Montr\'{e}al, QC H3A 0G4, Canada. (e-mail: shenghao@tju.edu.cn, xiaozhe.wang2@mcgill.ca)}
}

\markboth{Accepted by IEEE Transactions on Power Systems on January 27, 2019}%
{Sheng \MakeLowercase{\textit{et al.}}:
Probabilistic Power Flow Calculation using Non-intrusive Low-rank Approximation Method
}

\maketitle


\begin{abstract}
In this paper, a novel non-intrusive probabilistic power flow (PPF) analysis method based on the low-rank approximation (LRA) is proposed, which can accurately and efficiently estimate the probabilistic characteristics (e.g., mean, variance, probability density function) of the PPF solutions.
This method aims at building up a statistically-equivalent surrogate for the PPF solutions through a small number of power flow evaluations. By exploiting the retained tensor-product form of the univariate polynomial basis, a sequential correction-updating scheme is applied, making the total number of unknowns to be linear rather than exponential to the number of random inputs. 
Consequently, the LRA method is particularly promising for dealing with high-dimensional problems with a large number of random inputs. 
Numerical studies on the IEEE 39-bus, 118-bus, and 1354-bus systems show that the proposed method can achieve accurate probabilistic characteristics of the PPF solutions with much less computational effort compared to the Monte Carlo simulations. Even compared to the polynomial chaos expansion method, the LRA method can achieve comparable accuracy, while the LRA method is more capable of handling higher-dimensional problems. 
Moreover, numerical results reveal that the randomness brought about by the renewable energy resources and loads may inevitably affect the feasibility of dispatch/planning schemes.
\end{abstract}

\begin{IEEEkeywords}
Probabilistic power flow (PPF), Copula, low-rank approximation (LRA), Nataf transformation, polynomial chaos expansion (PCE).
\end{IEEEkeywords}

\IEEEpeerreviewmaketitle


\nomenclature[A]{LRA}{Low-rank approximation}%
\nomenclature[A]{MCS}{Monte Carlo simulation}%
\nomenclature[A]{PPF}{Probabilistic power flow}%
\nomenclature[A]{PCE}{Polynomial chaos expansion}%
\nomenclature[A]{PDF}{Probability distribution function}%
\nomenclature[A]{CDF}{Cumulative distribution function}%
\vspace{2ex}
\nomenclature[U]{$v_{i}$}{Wind speed $v$ in $m/s$ at bus $i$}%
\nomenclature[U]{$r_{i}$}{Solar radiation $r$ in $W/m^2$ at bus $i$}%
\nomenclature[U]{$P_{Li}$}{Real load at bus $i$}%
\nomenclature[U]{$U_{i}$}{The $i$-th random input, $U=[v,r,P_L]$}%
\vspace{2ex}
\nomenclature[F]{$G_{ij}$}{Real part of the entry $Y_{ij}$ in the admittance matrix}%
\nomenclature[F]{$B_{ij}$}{Imaginary part of the entry $Y_{ij}$ in the admittance matrix}%

\nomenclature[F]{$P_{Gi}$}{Real power from traditional generators at bus $i$}%
\nomenclature[F]{$Q_{Gi}$}{Reactive power from traditional generators at bus $i$}%
\nomenclature[F]{$P_{wi}$}{Real power from wind farms at bus $i$}%
\nomenclature[F]{$Q_{wi}$}{Reactive power from wind farms at bus $i$}%
\nomenclature[F]{$P_{pvi}$}{Real power from solar PV power plants at bus $i$}%
\vspace{2ex}
\nomenclature[L]{$\xi_{i}$}{Standard random variable related with input $U_i$}%
\nomenclature[L]{$\omega_{l}(\bm{\xi})$}{The $l$-th rank-one function}%
\nomenclature[L]{$b_{l}$}{Weighting factor of $\omega_{l}(\bm{\xi})$}%
\nomenclature[L]{$z_{k,l}^{(i)}$}{Coefficient of the $k$-th degree univariate polynomial}%
\nomenclature[L]{$\phi_{k}^{(i)}(\xi_{i})$}{The $k$-th degree univariate polynomial}%
\nomenclature[L]{$e_{r-1}$}{Approximation error of the $(r-1)$-th LRA surrogate}%
\nomenclature[L]{$\mathbb{E}$}{Expectation operator}%
\nomenclature[L]{$\mathbb{V}$}{Variance operator}%
\printnomenclature

%
\section{Introduction}

\IEEEPARstart{P}{ower} flow analysis is undoubtedly a fundamental and essential tool in power system operation and planning. It is employed to determine the system operating state, to compare different dispatching/planning schemes, and to provide the initial condition for other advanced applications. The growing integration of wind farms and utility-scale solar photo-voltaic (PV) power plants in transmission systems inevitably results in increasing degree of uncertainties to the operating state of a power system.
Therefore, adopting a probabilistic framework is essential to ensure realistic and accurate power flow analysis results.

Indeed, the proposal of probabilistic power flow (PPF) problem can be tracked back to 1970s, the aim of which is to incorporate uncertainties into the power flow problem \cite{BBorkowska74a}, \cite{RNAllan74a}. Many efforts were made thereafter. These methods can be generally divided into two categories: simulation methods and analytical methods. Monte Carlo simulations (MCS) and Monte-Carlo like simulations \cite{HYu09a}, \cite{JHuang11a}, \cite{MHajian13a} are the most widely used in the first category due to its simplicity, yet its computational effort grows exponentially as the random inputs increases.

Attempting to release the computational burden, some analytical methods are developed, which utilize mathematical approximations and assumptions. The convolution method proposed in \cite{RNAllan74a} applies the convolution to obtain the probability density function (PDF) of the power flow solutions. However, the computational burden is still high due to the time-consuming convolution.
Besides, the cumulant method was proposed in \cite{PZhang04a}, which can indirectly estimate the moments of responses.
Nevertheless, its accuracy will degrade if the random inputs have large variations. 
Another representative method is the point estimation method \cite{CLSu05a}, \cite{JMMorales07a}, which is computationally efficient yet its accuracy will deteriorate as the dimension of input random variables or the order of moments increases.
In addition, both the cumulant method and point estimation method can not provide PDF and cumulative distribution function (CDF) of the responses directly, and hence series expansions were proposed to work together with them including Gram-Charlier series \cite{PZhang04a}, Cornish-Fisher series \cite{FJRuizRodriguez12a}, and Edgeworth series \cite{MFan12a}. 
However, it has been shown in \cite{MFan12a} that the convergence of the series expansions can not always be guaranteed.

Another class of methods for uncertainty quantification is meta-modeling, which aims to build a statistically-equivalent analytical representation for the desired responses (e.g., the PPF solutions in this study) using a small number of model evaluations. Particularly, the polynomial chaos expansion (PCE) is the most popular one because (i) it has strong mathematical basis; (ii) it works with deterministic tools in a non-intrusive fashion; (iii) it can provide accurate and comprehensive statistical properties of responses with low computational effort.
The PCE method has been applied in the context of power systems to study the probabilistic power flow \cite{ZYRen16a}, \cite{FNi17a}, the load margin problem \cite{EHaesen09a}, and the available delivery capability problem \cite{HSheng18a}.

An emerging method, alternative to PCE, is the canonical low-rank approximations (LRA), which employs the canonical decomposition to express the desired response as a sum of rank-one functions \cite{KKonakli16a}. The original idea of canonical decomposition dates back to 1927 \cite{FHitchcock27a} and becomes an attractive method for uncertainty quantification of structural vibration problems recently \cite{ADoostan13a}-\cite{MChevreuil15a}.
By exploiting the tensor-product structure of the multivariate polynomial basis, LRA can provide polynomial representations in highly compressed formats.
The outstanding advantage of the LRA over the PCE is that the number of coefficients to be solved grows linearly rather than exponentially with the number of inputs, 
making it more powerful on dealing with high dimensional problems. 

In this paper, we apply the LRA approach to solve the probabilistic power flow problem in which various uncertainties from the wind power, the solar PV and the loads are incorporated. 
To the knowledge of authors, the paper seems to be \textbf{the first attempt} to apply the LRA method in the context of power system. 
The contributions of the paper are as below:
\begin{itemize}
\item A computationally efficient yet accurate algorithm is proposed to evaluate the PPF problem. 
Particularly, in the proposed LRA method:
\begin{itemize}
\item Random variables with diverse marginal distributions can be accommodated using proper polynomial basis. The correlations among them can be incorporated using the Nataf transformation.
\item A sequential correction-updating scheme is employed to build up the LRA model, making the total number of unknown coefficients to be linear rather than exponential to the number of random inputs. 
\item The mean and variance of the bus voltages and line flows can be obtained analytically without evaluating a large number of samples. 
\end{itemize}
\item Accurate probabilistic characteristics of the bus voltages and line flows can be achieved by the proposed methodology with much less computational effort compared to the Latin hypercube sampling (LHS)-based MCS.
\item A higher capability of dealing high-dimensional problems can be reached compared to the PCE method because the number of unknown coefficients grows only linearly with the number of random inputs.
\end{itemize}

It should be noted that a relevant problem "optimal power flow (OPF)", originating from the economic dispatch problem \cite{JCarpentier62a}, \cite{HWDommel68a} is fundamentally different from the power flow or PPF study considered in this paper. The power flow aims to determine the operating state, i.e., bus voltages and line flows, of the power system under given loads, generations and network conditions. It is formulated as a set of nonlinear equations. The PPF extends the power flow to assess the probability characteristics of the operating state for \textbf{a range of} loads, generations and network conditions. In contrast, the OPF aims to determine the \textbf{\textit{optimal}} operating state of the power system by simultaneously minimizing a given objective function and satisfying certain physical and operating constraints. It is typically formulated as an optimization problem. The probabilistic OPF \cite{MMadrigal98a} and chance-constrained OPF \cite{HZhang11a} are two relative terms of PPF in the probabilistic framework. They have been tailored for various applications in both transmission and distribution systems, e.g., decentralized energy trading \cite{SBahrami18a}. In the rest of the paper, we focus only on PPF. 

The remaining of the paper is organized as follows. Section II introduces the probabilistic power system models. 
Section III describes the mathematical formulation of PPF problem. Section IV elaborates the low-rank approximation method and its implementation in probabilistic power flow analysis. The detailed algorithm to assess the PPF is summarized in Section V. The simulation results on the modified IEEE 39-bus, 118-bus and 1354-bus systems are shown in Section VI. Conclusions and perspectives are given in Section VII.

%
\section{The Probabilistic Power System Model}
Power systems are subjected to various uncertainties and randomness from different sources, ranging from renewable generations, load variations, topology changes, to unexpected outages and faults. In this study, we focus on the randomness brought about by the renewable generators and loads. The other sources of uncertainty can also be integrated into the formulation if needed. Correspondingly, the random inputs of interests are wind speed ${v}$, solar radiation ${r}$ and load power ${P_{L}}$. Each random input typically can be expressed as a random variable associated with a probabilistic density function (PDF) ${X\sim f_{X}(x)}$. 

\subsection{Wind Generation}
In long-time scale, the wind speed in many locations around the world can be represented by a Weibull distribution \cite{SHKaraki99a}, \cite{Xiaozhe15}, \cite{Xiaozhe17}, 
the probability density function of which is

\small
\begin{equation}
\label{eq:wind_pdf_weibull}
f_{V}\left( v \right)=\frac{k}{c}{{\left( \frac{v}{c} \right)}^{k-1}} \exp \left[ -{{\left( \frac{v}{c} \right)}^{k}} \right]
\end{equation}
\normalsize
where $v$ is the wind speed, $k$ and $c$ are the shape and scale parameters, respectively.
Several methods have been proposed to extract the $k$ and $c$ from historical wind speed data, such as the moment method \cite{SHJangamshetti99a} and the maximum likelihood method \cite{SAAkdag09a}.
In short-time scale, however, the wind speed can be modeled as a Gaussian distribution \cite{XRan16a} in which the mean value is equal to the forecasted wind speed and the variance represents the forecasted error.

As a result, the active power output ${{P}_{w}}$ can be calculated by the piece-wise wind speed-power output relation \cite{SRoy02a}
\small
\begin{equation}
\label{eq:wind_p_val}
{{P}_{w}}(v)=\left\{ \begin{array}{*{35}{l}}
0 & v\le {{v}_{in}} \mbox{ or } v>{{v}_{out}} \\
\displaystyle \frac{v-{{v}_{in}}}{{{v}_{rated}}-{{v}_{in}}}{{P}_{r}} & {{v}_{in}}<v\le {{v}_{rated}} \\
{{P}_{r}} & {{v}_{rated}}<v\le {{v}_{out}} \\
\end{array} \right.
\end{equation}
\normalsize
where ${{v}_{in}}$, ${{v}_{out}}$ and ${{v}_{rated}}$ are the cut-in, cut-out, and rated wind speed ($m/s$), ${{P}_{r}}$ is the rated wind power ($kW$). 
Once the active power is obtained, 
the reactive power can be determined according to the speed control type of the wind turbine \cite{EHCamm09b} since the wind turbine can be modelled as either a constant P-Q bus or a constant P-V bus with given Q-limits. 

\subsection{Solar Generation}

Typically, the solar radiation in long time scale is represented by the Beta distribution \cite{SHKaraki99a} 
\small
\begin{equation}
\label{eq:solar_pdf_beta}
f_{R}\left( r \right)=\frac{\Gamma(\alpha+\beta)}{\Gamma(\alpha)\Gamma(\beta)}{{\left( \frac{r}{{{r}_{\max}}} \right)}^{\alpha-1}}{{\left( 1-\frac{r}{{{r}_{\max}}} \right)}^{\beta-1}}
\end{equation}
\normalsize
where $\alpha$ and $\beta$ are the shape parameters of the distribution, 
which can be fitted from historical solar radiation data by several methods, such as the moment method \cite{FYEttoumi02a} and the maximum likelihood method \cite{CBOwen08a}. 
$\Gamma$ denotes the Gamma function, $r$ and ${{r}_{\max}}$ (${W/m}^{2}$) are the respective actual and maximum solar radiations. Similar to the wind speed, the solar radiation in short-time scale can be modeled as a Gaussian distribution provided that accurate mean value is available \cite{XRan16a}.

The active power ${{P}_{pv}}$ corresponding to the solar radiation $r$ is determined by the following piece-wise function \cite{SHKaraki99a} 
\small
\begin{equation}
\label{eq:solar_p_val}
{{P}_{pv}}(r)=\left\{ \begin{array}{*{35}{l}}
\displaystyle \frac{{{r}^{2}}}{{{r}_{c}}{{r}_{std}}}{{P}_{r}} & 0\le r<{{r}_{c}} \\
\displaystyle \frac{r}{{{r}_{std}}}{{P}_{r}} & {{r}_{c}}\le r \le {{r}_{std}} \\
{{P}_{r}} & r>{{r}_{std}} \\
\end{array} \right.
\end{equation}
\normalsize
where ${{r}_{c}}$ is a certain radiation point set usually as 150 $W/m^2$, ${{r}_{std}}$ is the solar radiation in the standard environment, ${{P}_{r}}$ is the rated power of the solar PV. The reactive power ${{Q}_{pv}}$ is assumed to be zero in this study according to the solar generation is required to inject into the power grid at unit power factor \cite{WECC10}.
However, similar to the wind farm, the PV power plant can be modeled as P-Q bus or P-V bus with Q-limits in PPF once similar interconnection requirements are formulated and applied in the near future \cite{AEllis12a}.

\subsection{Load Variation}

By nature, load demand is uncertain in power systems. It is a common practice to model the load uncertainty by Gaussian distribution with specified mean value ${{\mu}_{P}}$ and variance ${{\sigma}_{P}}$ \cite{BBorkowska74a}, \cite{RBillinton08a}, which are typically provided by the load forecaster and historical data, respectively. Traditionally, only the active power is predicted, whereas the reactive power is calculated under the assumption of constant power factor \cite{WYLi13a}.

%
\section{Mathematical Formulation of Probabilistic Power Flow} \label{PPF_Formulation}
The power flow equations can be represented as
\small
\begin{equation}
\label{eq:power_flow}
{\bm{f}}(\bm{x})=\begin{bmatrix}
{P_{Gi}-P_{Li}-P_{i}(\bm{x})} \\
{Q_{Gi}-Q_{Li}-Q_{i}(\bm{x})} \end{bmatrix}=0
\end{equation}
\normalsize
\noindent where
\small
\begin{equation}
\label{eq:bus_flow_out}
\begin{gathered}
P_{i}(\bm{x}) = V_{i}\sum\limits_{j=1}^{N}{{V_{j}(G_{ij}\cos \theta_{ij}+B_{ij}\sin \theta_{ij})}} \\
Q_{i}(\bm{x}) = V_{i}\sum\limits_{j=1}^{N}{{V_{j}(G_{ij}\sin \theta_{ij}-B_{ij}\cos \theta_{ij})}}
\end{gathered}
\end{equation}
\normalsize
and ${\bm{x}}={{[{\theta},{V}]}^{T}}$, e.g., voltage angles and magnitudes for all buses; $G_{ij}$ and $B_{ij}$ are the real and imaginary part of the entry $Y_{ij}$ in the bus admittance matrix.

Let $v$, $r$ and ${{P}_{L}}$ be the random vectors that represent wind speeds, solar radiations and load variations, respectively, the probabilistic power flow (PPF) equations of a $N$-bus system can be described as below. Specifically, for P-Q type buses, the PPF equations are:
\small
\begin{equation}
\label{eq:pq_bus_pq}
\begin{gathered}
P_{Gi}+P_{wi}({{v}_{i}})+P_{pvi}({{r}_{i}})-P_{Li}(P_{Li})-P_{i}(\bm{x})=0 \\
Q_{Gi}+Q_{wi}({{v}_{i}})-Q_{Li}(P_{Li})-Q_{i}(\bm{x})=0
\end{gathered}
\end{equation}
\normalsize
\noindent For P-V type buses, the corresponding PPF equations are:
\small
\begin{equation}
\label{eq:pv_bus_pvq}
\begin{gathered}
P_{Gi}+P_{wi}({{v}_{i}})+P_{pvi}({{r}_{i}})-P_{Li}(P_{Li})-P_{i}(\bm{x})=0 \\
V_{i}={{V}_{i0}} \\
Q_{Gi}=-Q_{wi}({{v}_{i}})+Q_{Li}(P_{Li})+Q_{i}(\bm{x}) \\
{{Q}_{min,i}}\le Q_{Gi}\le {{Q}_{max,i}}
\end{gathered}
\end{equation}
\normalsize
\noindent where $P_{wi}({{v}_{i}})$, $P_{pvi}({{r}_{i}})$, $P_{Li}$ and $P_{Gi}$ are the real power injection from the wind turbine, the solar PV, the load, and the conventional generator at bus $i$; $Q_{wi}({{v}_{i}})$, $Q_{Li}$ and $Q_{Gi}$ are the reactive power injection from the wind power, the load, and the conventional generator at bus $i$. If $Q_{Gi}$ exceeds its limits, e,g., ${{Q}_{min,i}}$ or ${{Q}_{max,i}}$, then the terminal bus switches from P-V to P-Q with $Q_{Gi}$ fixed at the violated limit.

In fact, the set of PPF equations \eqref{eq:pq_bus_pq}-\eqref{eq:pv_bus_pvq} can be described in the following compact form:
\small
\begin{equation}
\label{eq:ppf_equation}
\begin{gathered}
{\bm{f}} \left( {\bm{x},\bm{U}} \right)=0 \\
\end{gathered}
\end{equation}
\normalsize
where ${\bm{x}}={{[{\theta},{V}]}^{T}}$ is the state vector, ${\bm{U}=\left[ \bm{v},\bm{r},\bm{{P}_{L}} \right]}$ is the random vector describing the wind speed, the solar radiation, and the load power.

%
\section{Canonical Low-rank Approximation using Polynomial Basis} \label{LRA_Section}
This section presents the general framework of low-rank approximation of a multivariate stochastic response function. 
For simplicity, we first consider a scalar response function of independent inputs, the case of dependent inputs will be addressed in Section \ref{LRA_Section}--E.

\subsection{Low-rank Approximation with Polynomial Basis}

Consider a random vector ${\bm{\xi}}$ = (${{\xi}_{1},{\xi}_{2},...,{\xi}_{n},}$) with joint probability density function (PDF) ${f_{\bm{\xi}}}$ and marginal distribution functions ${f_{\xi_{i}},i=1,...,n}$ (${\xi_{i}}$ is related with the random variables ${U_{i}}$ in \eqref{eq:ppf_equation}, see Section \ref{PPF_Formulation}), then the canonical rank-$r$ approximation \cite{KKonakli16a} of the target stochastic response (e.g., probabilistic bus voltages or branch flows in this study) ${Y = g(\bm{\xi})}$ can be represented by:

\small
\begin{equation}
\label{eq:lra_rank_R}
{Y}\approx{\hat{Y}}=\hat{g}\left( \bm{\xi} \right)=\sum\limits_{l=1}^{r}{{{b}_{l}}{{\omega}_{l}}\left( \bm{\xi} \right)}
\end{equation}
\normalsize
\noindent in which $b_{l},l=1,...,r$ are normalizing weighting factors, and ${{\omega}_{l}}$ is a rank-one function of $\bm{\xi}$ in the form of 
\small
\begin{equation}
\label{eq:lra_rank_one}
\omega_{l}(\bm{\xi})=\prod\limits_{i=1}^{n}{{{v}_{l}^{(i)}}( {{\xi}_{i}})}
\end{equation}
\normalsize

\noindent where ${v_{l}^{(i)}}$ denotes the ${i}$-th dimensional univariate function in the ${l}$-th rank-one function.
For most applications, the number ${r}$ of rank-one terms is usually small (under 5), hence \eqref{eq:lra_rank_R} and \eqref{eq:lra_rank_one} represent a canonical low-rank approximation.

In order to obtain the rank-$r$ approximation, a natural choice is expanding ${v_{l}^{(i)}}$ on a polynomial basis ${\{\phi_{k}^{(i)}, k \in \mathbb{N}\}}$ that is orthogonal to ${f_{\xi_{i}}}$, the resulting rank-$r$ approximation takes the form:
\small
\begin{equation}
\label{eq:lra_rank_r_pce}
{\hat{Y}}=\hat{g}(\bm{\xi})=\sum\limits_{l=1}^{r}{{{b}_{l}}\left[ \prod\limits_{i=1}^{n}{\left( \sum\limits_{k=0}^{{{p}_{i}}}{z_{k,l}^{\left( i \right)}\phi _{k}^{(i)}\left( {{\xi}_{i}} \right)} \right)} \right]}
\end{equation}
\normalsize
where ${\phi}_{k}^{(i)}$ denotes the $k$-th degree univariate polynomial in the $i$-th random input, $p_{i}$ is the maximum degree of ${\phi}^{(i)}$ and ${z}_{k,l}^{(i)}$ is the coefficient of ${\phi}_{k}^{(i)}$ in the $l$-th rank-one function.

Building the low-rank approximation for desired responses in \eqref{eq:lra_rank_r_pce} requires: (i) choose an appropriate univariate polynomial for each random input; (ii) solve the polynomial coefficients ${z_{k,l}^{(i)}}$ as well as the weighing factors ${b_{l}}$. 
This process relies on a set of samples of $\bm{\xi}$, which are usually termed as experimental design (ED) and their corresponding accurate responses $\bm{Y}$. 

To illustrate the construction of $r$-th rank-one function, consider a 3-dimensional input random vector ${\bm{\xi}=(\xi_{1},\xi_{2},\xi_{3})}$ with mutually independent components ${{{\xi}_{1}}\sim \mbox{Weibull}\left( \lambda,k \right)}$, ${{{\xi}_{2}}\sim \mbox{Beta}\left( \alpha,\beta \right)}$ and ${{{\xi}_{3}}\sim N\left( 0,1 \right)}$ respectively, then the $r$-th rank-one function $\omega_{r}$ that corresponds to ${p_{i}=2, i=1,2,3}$ is:
\small
\begin{equation}
\label{eq:lra_illustration}
\begin{gathered}
\omega_{r}(\xi)={{v}_{r}^{(1)}}({{\xi}_{1}})\times{{v}_{r}^{(2)}}({{\xi}_{2}})\times{{v}_{r}^{(3)}}({{\xi}_{3}}) \\
{{v}_{r}^{(1)}}({{\xi}_{1}})={z_{0,r}^{(1)}\phi_{0}^{(1)}({{\xi}_{1}})}+{z_{1,r}^{(1)}\phi_{1}^{(1)}({{\xi}_{1}})}+{z_{2,r}^{(1)}\phi_{2}^{(1)}({{\xi}_{1}})} \\
{{v}_{r}^{(2)}}({{\xi}_{2}})={z_{0,r}^{(2)}\phi_{0}^{(2)}({{\xi}_{2}})}+{z_{1,r}^{(2)}\phi_{1}^{(2)}({{\xi}_{2}})}+{z_{2,r}^{(2)}\phi_{2}^{(2)}({{\xi}_{2}})} \\
{{v}_{r}^{(3)}}({{\xi}_{3}})={z_{0,r}^{(3)}\phi_{0}^{(3)}({{\xi}_{3}})}+{z_{1,r}^{(3)}\phi_{1}^{(3)}({{\xi}_{3}})}+{z_{2,r}^{(3)}\phi_{2}^{(3)}({{\xi}_{3}})} \\
\end{gathered}
\end{equation}
\normalsize
where ${z_{0,r}^{(i)}}$, ${z_{1,r}^{(i)}}$ and ${z_{2,r}^{(i)}}$ are the coefficients corresponding to the zero-, first- and second-order univariate polynomial ${\phi_{0}^{(i)}({{\xi}_{i}})}$, ${\phi_{1}^{(i)}({{\xi}_{i}})}$ and ${\phi_{2}^{(i)}({{\xi}_{i}})}$ respectively. Now we have the $r$-th rank-one function ${\omega_{r}(\xi)}$ with known polynomial basis, the next step is to determine the coefficients $\{{z_{0,r}^{(i)}},{z_{1,r}^{(i)}}, {z_{2,r}^{(i)}} : i=1,2,3\}$.

\subsection{Selection of the Univariate Polynomial Basis}

To reduce the computational effort for building an accurate low-rank approximation \eqref{eq:lra_rank_r_pce} for each response of interests, it is crucial to choose a proper polynomial ${\phi}_{i}$ for the $i$-th random input ${\xi}_{i},i=1,2,...,n$; otherwise, lower convergence rate and higher degree of polynomial basis may be needed \cite{DBXiu02a}, which will hamper the capability of LRA in dealing with high-dimensional problems.

Table \ref{tab:gpce_mapping} shows a set of typical continuous distributions and the respective optimal Wiener-Askey polynomial basis \cite{DBXiu02a}, which can ensure the exponential convergence rate. In case that $f_{\xi_{i}}$ is out of the list in Table \ref{tab:gpce_mapping}, two options are available: (i) employ the isoprobabilistic transformation \cite{RLebrun09b} to project $\xi_{i}$ to a typical distribution in Table \ref{tab:gpce_mapping}, and then choose its corresponding polynomial basis; (ii) adopt the discretized Stieltjes procedure to numerically construct a set of univariate orthogonal polynomial basis in the form: ${\tilde{\pi}=\frac{{{\pi}_{k}}}{\left\langle {{\pi}_{k}},{{\pi}_{k}} \right\rangle}}$ for $\xi_{i}$. In this paper, we carry out the second option to recurrently compute the orthogonal polynomial basis \cite{SMarelli17a}:

\begin{table}[t]
\renewcommand{\arraystretch}{1.3}
\caption{Standard Forms of Classical Continuous Distributions and their Corresponding Orthogonal Polynomials \cite{DBXiu02a}}
\label{tab:gpce_mapping}
\begin{center}
\begin{tabular}{c|c|>{\centering}p{1.5cm}|c}
\hline
\bfseries Distribution & \bfseries Density Function & \bfseries Polynomial & \bfseries Support \\
\hline
Normal & ${\frac{1}{\sqrt{2\pi}}{{e}^{-{{x}^{2}}/2}}}$ & Hermite & (-$\infty$,$\infty$) \\
\hline
Uniform & ${\frac{1}{2}}$ & Legendre & [-1,1] \\
\hline
Beta & ${\frac{{{(1-x)}^{\alpha}}{{(1+x)}^{\beta}}}{{{2}^{\alpha+\beta+1}}{B(\alpha+1,\beta+1)}}}$ & Jacobi & [-1,1] \\
\hline
Exponential & ${{e}^{-x}}$ & Laguerre & (0,$\infty$) \\
\hline
Gamma & ${\frac{{{x}^{\alpha}}{{e}^{-x}}}{\Gamma(\alpha+1)}}$ & Generalized Laguerre & [0,$\infty$) \\
\hline
\end{tabular}
\end{center}
\begin{tablenotes}
\item * The Beta function is defined as $B(p,q)=\frac{\Gamma(p)\Gamma(q)}{\Gamma(p+q)}$.
\end{tablenotes}
\end{table}
\small
\begin{equation}
\label{eq:arbi_poly_recur}
\begin{gathered}
{\sqrt{{b}_{k+1}}}{{\tilde{\pi}}_{k+1}}\left( {\xi_{i}} \right)=\left( {\xi_{i}}-{{d}_{k}} \right){{\tilde{\pi}}_{k}}\left( {\xi_{i}} \right)-\sqrt{{b}_{k}}{{\tilde{\pi}}_{k-1}}\left( {\xi_{i}} \right) \\
{{{{\tilde{\pi}}_{-1}}\left( {\xi_{i}} \right)=0,\tilde{\pi}}_{0}}\left( {\xi_{i}} \right)=1
\end{gathered}
\end{equation}
\normalsize
\small
\begin{equation}
\label{eq:arbi_poly_recur_coef}
\begin{gathered}
{{d}_{k}}=\frac{\left\langle \xi {{\pi}_{k}},{{\pi}_{k}} \right\rangle }{\left\langle {{\pi}_{k}},{{\pi}_{k}} \right\rangle }, k \ge 0 \\
{{b}_{0}}=\left\langle {{\pi}_{0}},{{\pi}_{0}} \right\rangle ,{{b}_{k}}=\frac{\left\langle {{\pi}_{k}},{{\pi}_{k}} \right\rangle }{\left\langle {{\pi}_{k-1}},{{\pi}_{k-1}} \right\rangle }, k \ge 1
\end{gathered}
\end{equation}
\normalsize

\noindent where $k$ denotes the polynomial degree. 

It is worth mentioning that the fundamental premise of applying polynomial basis in Table \ref{tab:gpce_mapping} or equations \eqref{eq:arbi_poly_recur}-\eqref{eq:arbi_poly_recur_coef} is that the PDF of random inputs are known in advance. This assumption is reasonable if abundant historical data (several years) of wind speed and solar radiation are available, from which the wind speed and solar radiation can be well estimated 
(for wind \cite{SHJangamshetti99a}, \cite{SAAkdag09a}; for solar \cite{ZMSalameh95a}, \cite{FYEttoumi02a}). In case the historical data cannot be well fitted by any typical distribution, the data-driven polynomial chaos method offers an alternative choice to construct the polynomial basis from dataset \cite{SOladyshkin12a}.

Now we have chosen an optimal polynomial basis for each random input. The next step is to determine the coefficients ${z_{k,l}^{(i)}}$ as well as the weighing factors ${b_{l}}$ in \eqref{eq:lra_rank_r_pce}.

\subsection{Calculation of the Coefficients and Weighing Factors}\label{section_solve_lra}

Based on a experiment design of size $M_{C}$, i.e., a set of samples of $\bm{\xi}_{C}=\{\bm{\xi}^{(1)},\bm{\xi}^{(2)},...,\bm{\xi}^{(M_{C})}\}$ and corresponding responses $\bm{y}_{C}=\{y^{(1)},y^{(2)},...,y^{(M_{C})}\}$ which evaluated by deterministic tools, different algorithms have been proposed in the literature for solving the LRA coefficients and weighing factors in a non-intrusive manner \cite{ADoostan13a}, \cite{MChevreuil15a}, \cite{PRai14a}.
The sequential correction-updating scheme (Algorithm \ref{algo:scheme}) presented in \cite{MChevreuil15a} is employed in this study due to its efficiency and capability of constructing low-rank approximation using less sample evaluations. 
In the ${r}$-th correction step, a new rank-one function ${\omega_{r}}$ is built, while in the ${r}$-th updating step, the set of weighing factors ${\{b_{1},...,b_{r}\}}$ is determined. This process continues until the applied error index stop decreasing \cite{MChevreuil15a}.

\textbf{Correction step}:
the ${r}$-th correction step aims to find a new rank-one tensor $\omega_{r}$, which can be obtained by solving the following minimization problem:
\small
\begin{equation}
\label{eq:lra_rank_one_min}
\begin{aligned}
{{\omega}_{r}}(\bm{\xi}) &= \arg \underset{\omega \in W}{\mathop{\min}}\,\left\| {{e}_{r-1}}-\omega \right\|_{\bm{\xi}_{C}}^{2} \\
& = \arg \underset{\omega \in W}{\mathop{\min}}\,\sum_{m=1}^{M_{C}}{\left[y^{(m)}-\hat{g}_{r-1}(\bm{\xi}^{(m)})-\omega(\bm{\xi}^{(m)})\right]^{2}}
\end{aligned}
\end{equation}
\normalsize
\noindent where ${W}$ represents the space of rank-one tensors, ${{e}_{r-1}}=(g-{{\hat{g}}_{r-1}})$ is the approximation error of the response $Y$ at the ${(r-1)}$-th step, $\|.\|^{2}$ represents the norm 2 of the residual after the new rank-one tensor $w$ is applied, and the subscript ${\bm{\xi}_{C}}$ indicates that the minimization is carried over the whole set of samples in the experiment design $(\bm{\xi}_{C},\bm{y}_{C})$.

By exploiting the retained tensor-product form of the univariate polynomial basis, as shown in \eqref{eq:lra_rank_r_pce},
typical scheme for solving equation \eqref{eq:lra_rank_one_min} is the alternated least-square (ALS) minimization, which involves sequential minimization along each dimension ${i=1,...,n}$ to solve the corresponding polynomial coefficients ${\bm{z}_{r}^{(i)}=(z_{0,r}^{(i)},...,z_{p_{i},r}^{(i)})}$.
The total number of coefficients to be solved in each correction step is $\sum\nolimits_{i=1}^{n}{({p}_{i}+1)}$, which grows linearly as the number of random inputs $n$ increases.
Since ${{\omega }_{r}}$ is the product of $v_{r}^{(i)}({\xi}_{i})$ as shown in \eqref{eq:lra_rank_one}, $v_{r}^{(i)}({\xi}_{i})$ can be initialized as $1.0$.

In the minimization along the ${i}$-th dimension, the polynomial coefficients corresponding to all other dimensions are "frozen" at their current values and the coefficients ${\bm{z}_{r}^{(i)}=(z_{0,r}^{(i)},...,z_{p_{i},r}^{(i)})}$ can be determined by:
\small
\begin{equation}
\label{eq:lra_coeff}
\bm{z}_{r}^{\left( i \right)}=\arg \underset{\bm{\zeta} \in {{R}^{({{p}_{i}}+1)}}}{\mathop{\min}}\,\left\| {{e}_{r-1}}-C_{i}\left( \sum\limits_{k=0}^{{{p}_{i}}}{{{\zeta}_{k}}\phi _{k}^{\left( i\right)}} \right)\right\|_{\bm{\xi}_{C}}^{2}
\end{equation}
\normalsize

\noindent where $C_{i}$ is a scalar
\small
\begin{equation}
\label{eq:lra_coeff_c}
C_{i}=\prod\limits_{j\ne i}{v_{r}^{(j)}(\xi_{j})}=\prod\limits_{j\ne i}{\left( \sum\limits_{k=0}^{{{p}_{j}}}{z_{k,r}^{\left( j \right)}\phi _{k}^{\left( j \right)}\left( {{\xi}_{j}} \right)} \right)}
\end{equation}
\normalsize

\textbf{Updating step}: After the ${r}$-th correction step is completed, the algorithm proceed to the ${r}$-th updating step to determine the weighing factor $b_{r}$ of the newly solved rank-one function ${{\omega }_{r}(\bm{\xi})}$, meanwhile, the set of existing weighing factors ${\bm{b}=(b_{1},...,b_{r-1})}$ 
are updated too. The updating step can be achieved by solving the following minimization problem
\small
\begin{equation}
\label{eq:lra_weight}
\bm{b}=\arg \underset{\bm{\beta} \in {{R}^{r}}}{\mathop{\min}}\,\left\| g-\sum\limits_{l=1}^{r}{{{\beta}_{l}}{{\omega}_{l}}} \right\|_{\bm{\xi}_{C}}^{2}
\end{equation}
\normalsize

\textbf{Stop criteria}: The correction-updating scheme successively adds new rank-one function to improve the accuracy of the approximation \eqref{eq:lra_rank_r_pce}, hence, error reduction in two successive iterations becomes an natural stop criteria for this process. In this study, the relative empirical error is employed which is given by:
\small
\begin{equation}
\label{eq:lra_error}
{{\hat{e}}_{r}}=\frac{\left\| {{e}_{r-1}}-{{\omega}_{r}} \right\|_{\bm{\xi}_{C}}^{2}}{\mathbb{V}(\bm{y}_{C})}
\end{equation}
\normalsize
where $\mathbb{V}(\bm{y}_{C})$ denotes the empirical variance of the desired response over the experimental design.
\begin{algorithm}
\caption{The sequential correction-updating scheme to calculate LAR coefficients and weighting factors}
\label{algo:scheme}
\textcolor{black}{\KwIn{${\bm{\xi}_{C}}$, ${\bm{y}_{C}}$, ${p_{i}}$, ${\bm{\phi}^{(i)}=\{{\phi}_{0}^{(i)},...,\bm{\phi}_{p_i}^{(i)}\}}$ for $i=1,...,n$}
\KwOut{$g_{r}(\bm{\xi})$, $r$, $\bm{b}=(b_1,...,b_r)$, $\bm{z}_{l}^{(i)}=({z}_{l,0}^{(i)},...,{z}_{l,p_{i}}^{(i)})$ for $i=1,...,n$, $l=1,...,r$}
\textbf{Initialize} $keep \gets 1$, $r \gets 0$, $\hat{g}_{0}(\bm{\xi}_{C})=0$ \\
\While{$keep>0$}{
    $r \gets r+1$ \\
    $e_{r-1} \gets \bm{y}_{C}-\hat{g}_{r-1}(\bm{\xi}_{C})$ \\
    \textbf{Set} $v_{r}^{(i)} \gets 1$ for $i=1,...,n$ \\
    \textbf{Correction step:} Solve coefficients along each input \\
    \For{$i \gets 1$ \textbf{to} $n$}{
        Calculate $C_{i}$ by (18) with solved $v_{r}^{(q)}$ for $q<i$; \\
        Solve (17) to obtain $\bm{z}_{r}^{(i)}$ for the $\bm{\phi}^{(i)}$. \\
        Update $v_{r}^{(i)}$ using solved $\bm{z}_{r}^{(i)}$. \\
    }
    \textbf{Updating step:} Solve (19) to obtain $b_{r}$ and updating $b_1,...,b_{r-1}$ as well. \\
    \textbf{Stop criteria:} Calculate the error index $\hat{e}_{r}$ using (20). \\
    \uIf{$\hat{e}_{r}>\hat{e}_{r-1}$}{
        $keep \gets 0$, $r \gets r-1$\;
    }\Else{
        Store the rank $r$ LRA $\hat{g}_{r}(\bm{\xi}) \gets \sum_{l=1}^{r}{{b}_{l}{\omega}_{l}}$
    }
}
\Return\;}
\end{algorithm}

Remark: (i) It is worth pointing out that the minimization problems in (\ref{eq:lra_coeff}) and (\ref{eq:lra_weight}) can be efficiently solved with the ordinary least-squares (OLS) technique because the dimension of unknowns are small.
(ii) When the LRA \eqref{eq:lra_rank_r_pce} for the desired response is built up, the response of any new samples can be evaluated efficiently by directly substituting to \eqref{eq:lra_rank_r_pce} instead of solving the original complex problem (e.g., the power flow problem \eqref{eq:ppf_equation}).

\subsection{Selection of optimal rank and polynomial degree}
Currently, there is no systematic way to identify the optimal rank $r$ and the polynomial degree $p_i$ for individual random input beforehand. In this study, we specify a candidate set, say ${\{1, 2, 3, 4, 5\}}$, for the rank and another one ${\{2, 3, 4, 5\}}$ for the polynomial degree, and assume the same degree for all the univariate polynomials. The rank selection is performed by progressively increasing the rank and applying the corrected error to select the best one. The selection of optimal degree can be implemented in a similar way.
Further study on how to determine an optimal rank and the degree in a systematic and efficient way is desired.

\subsection{Integration of Dependent Random Inputs}

So far, the random inputs are assumed to be mutually independent as requested by the LRA method.
To accommodate dependent random inputs with correlation matrix $\bm{\rho}$, the Nataf transformation \cite{RLebrun09b}, \cite{RLebrun09a} and the isoprobabilistic transformation can be employed to build up a mapping between $\bm{U}$ and the independent standard random variables $\bm{\xi}$: $\bm{u}=T(\bm{\xi})$, where $T$ is invertible. Therefore, the set of samples of ${\bm{\xi}}$ can be transformed back into samples of ${\bm{U}}$ to evaluate the corresponding responses ${\bm{Y}}$, after which the desired response $\bm{y}=g(T^{-1}(\bm{u}))$ can be expanded onto the polynomial basis with $\bm{\xi}$ using the aforementioned method \cite{KKonakli16a}.

\subsection{Moments of a Low-Rank Approximation}

Due to the orthogonality of the univariate polynomials that form the LRA basis (see (\ref{eq:lra_rank_r_pce})), the mean and variance of the meta-model can be obtained \textit{analytically} in terms of the polynomial coefficients and the weighing factors. In particular, the mean and variance of the LRA response are given respectively by \cite{KKonakli16a}:
\small
\begin{equation}
\label{eq:lra_post_mean}
{{\mu }_{y}}=\mathbb{E}\left[ {\hat{g}}\left( \bm{\xi} \right) \right]=\sum\limits_{l=1}^{r}{{{b}_{l}}\left( \prod\limits_{i=1}^{n}{z_{0,l}^{\left( i \right)}} \right)}
\end{equation}
\normalsize
\noindent and
\small
\begin{equation}
\label{eq:lra_post_var}
\sigma _{y}^{2}=\sum\limits_{l=1}^{r}{\sum\limits_{m=1}^{r}{{{b}_{l}}{{b}_{m}}\prod\limits_{i=1}^{n}{\left[ \left( \sum\limits_{k=0}^{{{p}_{i}}}{z_{k,l}^{\left( i \right)}z_{k,m}^{\left( i \right)}} \right)-z_{0,l}^{\left( i \right)}z_{0,m}^{\left( i \right)} \right]}}}
\end{equation}
\normalsize

Hence, if only mean and variance are of interests, \eqref{eq:lra_post_mean} and \eqref{eq:lra_post_var} can be applied directly without evaluating a large size of samples which is required by most of simulation-based methods. If other statistical properties (e.g., higher-order moments, PDF, CDF, etc.) are needed, one can sample the random inputs extensively and apply the solved functional approximation \eqref{eq:lra_rank_r_pce} to evaluate the corresponding response $\bm{y}$, from which the the statistics of interest can be obtained.

%
\section{Computation of Probabilistic Power Flow}
\begin{figure}[h]
\centering
\includegraphics[width=0.49\textwidth,trim={2.5cm 1.0cm 3.0cm 0},clip]{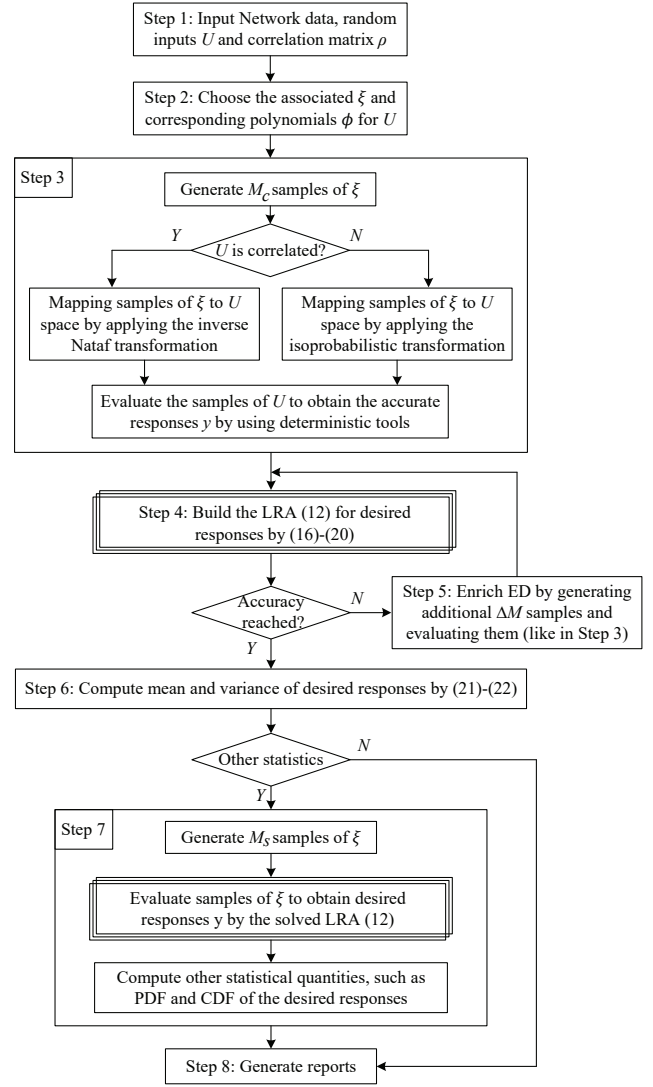}
\caption{The flow char of LRA-PPF method.}
\label{fig:lra_flowchart}
\end{figure}

In this section, a step-by-step description of the proposed probabilistic power flow calculation is summarized below, which is also illustrated in Fig. \ref{fig:lra_flowchart}:

\noindent Step 1: Input the network data, the probability distribution and the parameters of the random inputs $\bm{U}$, i.e., the wind speed, the solar radiation, the load active power, and their correlation matrix ${\bm{\rho}}$.

\noindent Step 2: Choose the independent standard variable ${\xi_{i}}$ and the corresponding univariate polynomial ${\phi_{i}}$ for each random input $U_{i}$.

\noindent Step 3: Generate an experimental design of size ${{M}_{C}}$: 
\begin{itemize}
\item i) Generate ${M_{C}}$ samples ${\bm{\xi}_{C}=(\xi^{(1)},\xi^{(2)},...,\xi^{({M_{C}})})}$ in the standard space by the LHS.
\item ii) Transform ${\bm{\xi}_{C}}$ into the physical space by the inverse Nataf transformation ${\bm{u}_C}=T^{-1}(\bm{\xi}_C)$.
\item iii) Evaluate the accurate responses (bus voltages and line flows in this study) ${\bm{y}_{C}=(y^{(1)},y^{(2)},...,y^{({M_{C}})})}$ of ${\bm{u}_{C}}$ by the deterministic power flow solver. Pass the sample-response pairs $({\bm{\xi}_{C}},{\bm{y}_{C}})$ to Step 4.
\end{itemize}

\noindent Step 4: Apply the algorithm in Section \ref{section_solve_lra} (Algorithm \ref{algo:scheme}) to solve the coefficients and the weighting factors of LRA to build the low-rank approximation (\ref{eq:lra_rank_r_pce}) for all desired responses $\bm{Y}$. If LRA for all responses have reached the prescribed accuracy, go to Step 6; otherwise, go to Step 5.

\noindent Step 5: Generate additional ${\Delta M_{C}}$ new samples and evaluate them, then go back to Step 4 using the enriched experiment design $({\bm{\xi}_{C}}+{\Delta \bm{\xi}_{C}},{\bm{y}_{C}}+{\Delta \bm{y}_{C}})$.

\noindent Step 6: Calculate the mean and variance of all responses through \eqref{eq:lra_post_mean} and \eqref{eq:lra_post_var}, respectively. If other statistical properties (e.g., higher-order moments, PDF, CDF, etc.) are needed, go to Step 7; otherwise, go to Step 8.

\noindent Step 7: Sample $\bm{\xi}$ extensively, e.g., ${{M}_{S}}$ samples, and apply the solved functional approximation (\ref{eq:lra_rank_r_pce}) to evaluate the corresponding responses $\bm{y}$ for all these samples. Compute the statistics of interests for each response. 

\noindent Step 8. Generate the result report.

Remark: the number of samples ${{M}_{C}}$ in Step 3 is usually much smaller than ${{M}_{S}}$ in Step 6. Unlike MCS, LRA does not solve power flow equations for all ${{M}_{S}}$ samples in Step 6, and hence it is much more efficient. The main computational effort of LRA lies in Step 3.

%

%
\section{Numerical Studies}

In this section, we apply the proposed LRA method to investigate the probabilistic power flow of the modified 39-bus, 118-bus, and 1354-bus systems \cite{MATPOWER}. The LHS-based Monte Carlos simulation serves as a benchmark for validating the accuracy and the performance of the proposed method.
In addition, a comparison between the LRA and the sparse PCE using the UQLab \cite{SMarelli14a} is also presented. 
In order to compare the accuracy of them, the error indices introduced in \cite{JMMorales07a} are adopted to indicate the distribution accuracy of responses as follows: 
\small
\begin{equation}
\label{eq:accuracy_comp}
\begin{gathered}
{\frac{\Delta \mu_{*}}{\mu_{mcs}}} \% = \left|\frac{\mu_{*}-\mu_{mcs}}{\mu_{mcs}} \right| \times 100\% \\
{\frac{\Delta \sigma_{*}}{\sigma_{mcs}}} \% = \left|\frac{\sigma_{*}-\sigma_{mcs}}{\sigma_{mcs}} \right| \times 100\% 
\end{gathered}
\end{equation}
\normalsize
where $\mu$ and $\sigma$ denote the mean and variance of the response, respectively. The star * represents the applied method, e.g., 'lra' for the LRA method and 'pce' for the sparse PCE method.
It should be noted that the same set of samples was used for building up the LRA and the sparse PCE in all simulation cases. 

In this study, we assume that the probability distributions and the associated parameters of all random inputs are available from up-front modeling. Particularly, the wind speed follows Weibull distribution, the parameters of which are $k=9.0$ and $c=2.15$; the solar radiation follows Beta distribution, the parameters of which are $\alpha=0.9$, $\beta=0.9$, $r=0$ and $s=1000$; the load power follows Normal distribution. For each individual load, the mean value is set to be its base case value and the variance is equal to 5\% of its mean value. The univariate polynomial basis used in \eqref{eq:lra_rank_r_pce} for Weibull, Beta and Normal distributions are chosen to be the numerical polynomial, the Jacobi polynomial and the Hermite polynomial, respectively, according to Table \ref{tab:gpce_mapping}.
It should be noted that the proposed method is not limited to any specific probability distribution. 
For simplicity, all wind generators share the same set of parameters $v_{rate}=15.0 \mbox{ m/s}$, $v_{in}=4.0 \mbox{ m/s}$, and $v_{out}=25.0 \mbox{ m/s}$. Similarly, all solar PVs have the same values $r_{c}=150.0 \mbox{ W/$m^2$}$, and $r_{std}=1000.0\mbox{ W/$m^2$}$, respectively. The linear correlation coefficient ${\rho_{ij}}$ between component $i$ and $j$ of wind speed $v$, solar radiations $r$ and load power $P_{L}$ are 0.5053, 0.8040, 0.4000, respectively.

\subsection{The Modified 39-Bus System}

The 39-bus system is a simplified New-England network which contains 10 generators, 21 loads and 47 branches. The total load of this network is 6254.23 MW and 1387.10 Mvar. In order to assess the impact of uncertainties on the feasibility of power flow solution of this network, 4 solar PV power plants of 120 MW are connected to bus $\{36, 37, 38, 39\}$ and 4 wind farms of 180 MW are connected to bus $\{32, 33, 34, 35\}$. There are totally 29 random inputs including 21 random loads. The renewable capacity penetration level is 17.44\%. 

We first apply the power flow tool to the deterministic system (i.e., without uncertainty), and the power flow solution is shown to be feasible, i.e., there is no voltage violation or thermal violation in the network. However, the system is prone to thermal violation, since we have scaled up the load power at bus $\{1, 3, 4, 15, 16, 18, 20, 21, 23, 24, 25, 26, 27, 28, 29\}$ by 10\%. In other words, we consider a heavy loading condition, under which the randomness may affect the static security level of the system as shown subsequently. 

\begin{figure}[h]
\centering
\includegraphics[width=0.45\textwidth,keepaspectratio=true,angle=0]{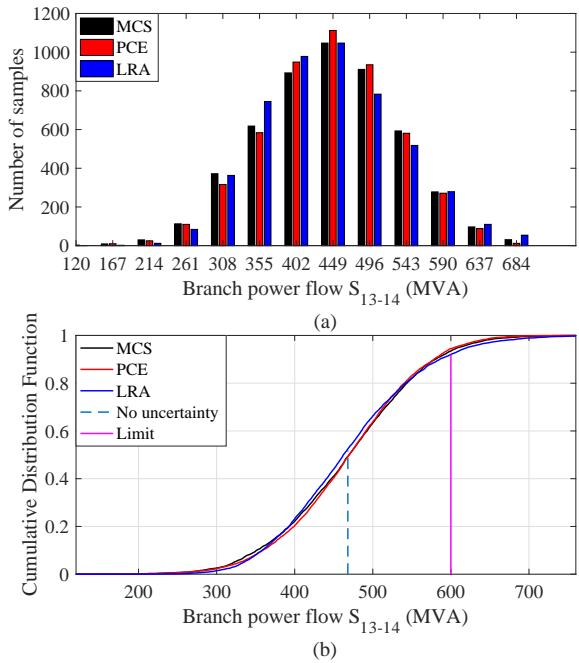}
\caption{The distribution of the branch flow $S_{13-14}$ computed by the MCS, the PCE and the LRA. They are almost overlapped. There is 8\% probability that the branch flow will be out of the allowed thermal limits due to the uncertainties. (a) Probability distribution of $S_{13-14}$. (b) Cumulative distribution function of $S_{13-14}$.}
\label{fig:branch_s_dist}
\end{figure}

Next, we exploit the proposed LRA method to assess the probabilistic characteristics (e.g., mean, variance, PDF and CDF) of these branch flows and compare the results with those of the LHS-based MCS and with those of the sparse PCE method.
Applying the proposed algorithm, 146 simulations are required in Step 4-5 to build up the LRAs \eqref{eq:lra_rank_r_pce} of the desired responses (i.e., the bus voltages, branch flows and generator reactive power), which consist of 1 rank-one functions with the highest polynomial degrees $p_{i}=2$. Once the coefficients and the weighting factors are computed, the mean and the standard deviation of the responses are computed in Step 6 and are compared with those of the LHS-based MCS and with those of the sparse PCE method as shown in Table \ref{tab:comp_mean_39} and \ref{tab:comp_var_39}. Furthermore, 5000 samples are generated in Step 7 to assess the PDF and the CDF of the responses. Fig. \ref{fig:branch_s_dist} shows the PDF and CDF of the branch flow $S_{13-14}$ computed by the the LHS-based MCS, the sparse PCE and the solved LRA, respectively. These results clearly demonstrate that the LRA method can provide accurate estimation for the probabilistic characteristics of the PPF solutions. Particularly, the accuracy of the LRA is comparable to that of the sparse PCE method.
\begin{table}[h]
\renewcommand{\arraystretch}{1.3}
\caption{Comparison of the estimated mean of the bus voltage, branch flow and generator reactive power}
\label{tab:comp_mean_39}
\centering
\begin{tabular}{c|c|c|c|c|c}
\hline
$V_{im}/S_{ij}$ & $\mu_{mcs}$ & $\mu_{pce}$ & $\mu_{lra}$ & ${\frac{\Delta \mu_{pce}}{\mu_{mcs}}} \%$ & ${\frac{\Delta \mu_{lra}}{\mu_{mcs}}} \%$ \\
%
%
\hline
$V_{8,m}$ & 0.9804 & 0.9801 & 0.9802 & -0.0322 & -0.0241 \\	
\hline
$V_{7,m}$ & 0.9805 & 0.9802 & 0.9803 & -0.0333 & -0.0248 \\
\hline
$S_{6-11}$ & 2.2189 & 2.2247 & 2.2436 & 0.2627 & 1.1129 \\
\hline
$S_{4-5}$ & 4.2638 & 4.2692 & 4.2716 & 0.1268 & 0.1824 \\
\hline
$S_{10-13}$ & 4.6271 & 4.6324 & 4.6298 & 0.1145 & 0.0577 \\
\hline
$S_{13-14}$ & 4.6821 & 4.6892 & 4.6853 & 0.1510 & 0.0670 \\
\hline
$Q_{32}$ & 2.7018 & 2.7186 & 2.7078 & 0.6214 & 0.2213 \\
\hline
$Q_{36}$ & 1.1593 & 1.1584 & 1.1605 & -0.0757 & 0.1062 \\
\hline
\end{tabular}
\begin{tablenotes}
\item * $\mu_{mcs}$, $\mu_{pce}$ and $\mu_{lra}$ represent the mean value of bus voltage or branch flow computed by the MCS, PCE, and LRA, respectively.
\end{tablenotes}
\end{table}
\begin{table}[h]
\renewcommand{\arraystretch}{1.3}
\caption{Comparison of the estimated standard deviation of the bus voltage, branch flow and generator reactive power}
\label{tab:comp_var_39}
\centering
\begin{tabular}{c|c|c|c|c|c}
\hline
${V_{im}/S_{ij}}$ & ${\sigma_{mcs}}$ & ${\sigma_{pce}}$ & ${\sigma_{lra}}$ & ${\frac{\Delta \sigma_{pce}}{\sigma_{mcs}} \%}$ & ${\frac{\Delta \sigma_{lra}}{\sigma_{mcs}} \%}$ \\
%
%
\hline
$V_{8,m}$ & 0.0155 & 0.0141 & 0.0154 & -8.9862 & -0.5056 \\
\hline
$V_{7,m}$ & 0.0158 & 0.0144 & 0.0157 & -9.0738 & -0.5686 \\
\hline
$S_{6-11}$ & 0.8112 & 0.6962 & 0.8210 & -14.1836 & 1.1999 \\
\hline
$S_{4-5}$ & 1.4085 & 1.3331 & 1.4131 & -5.3578 & 0.3225 \\
\hline
$S_{10-13}$ & 0.7968 & 0.7652 & 0.8034 & -3.9612 & 0.8265 \\
\hline
$S_{13-14}$ & 0.8835 & 0.8446 & 0.8899 & -4.4042 & 0.7202 \\
\hline
$Q_{32}$ & 0.4899 & 0.4321 & 0.4859 & -11.8063 & -0.8291 \\
\hline
$Q_{36}$ & 0.1057 & 0.0956 & 0.1043 & -9.5410 & -1.3486	\\
\hline
\end{tabular}
\begin{tablenotes}
\item * ${\sigma_{mcs}}$, ${\sigma_{pce}}$ and ${\sigma_{lra}}$ represent the standard deviation of bus voltage or branch flow computed by the MCS, PCE, and LRA, respectively.
\end{tablenotes}
\end{table}

Once we have the statistics of the quantities of interest, say the bus voltages and branch flows, one natural step afterwards is to see how the uncertainty actually affect the operating point of a power system. As shown by the CDF curve in Fig. \ref{fig:branch_s_dist}, the branch flow $S_{13-14}$ in the deterministic system (454MW, dot line) is far away from the thermal limit (600 MW, pink line). However, the probability corresponding to the thermal limit 600 MW (the pink line) is 92\%, indicating that there is still 8\% probability that the corresponding power flow solution will exceed the thermal limit when the uncertainties are considered. 

\begin{figure}[h]
\centering
\includegraphics[width=0.45\textwidth,keepaspectratio=true,angle=0]{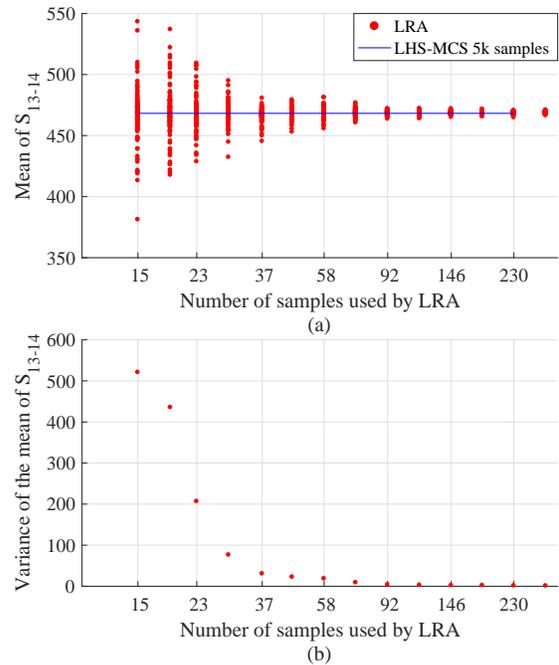}
\caption{The distribution of the mean of $S_{13-14}$ as the number of samples increases. Little improvement can be achieved by increasing the sample size after it reaches 5 times of the number of random inputs (i.e., 146 in this case). (a) Distribution of the mean of $S_{13-14}$. (b) Variance of the mean of $S_{13-14}$.}
\label{fig:branch_s_cr}
\end{figure}

In addition, the convergence rate of the LRA method is traced by increasing the sample size from $0.25n$ to $10n$ ($n$ is the number of random inputs) to verify the robustness of the proposed method. For each sample size, 100 replications are generated and are evaluated by the proposed LRA method. As shown in Fig. \ref{fig:branch_s_cr}, the statistics of the branch flow settle down to the values computed by the LHS-based MCS using 5000 samples when the sample size reaches $5n$ 
(i.e., 146 in this case), after which little improvement can be achieved by increasing the sample size. Similar results have been observed in other PPF solutions, indicating a nice property of the LRA that its simulation time (the required sample evaluations) grows linearly as the number of the random inputs increases.
\subsection{The Modified IEEE 118-Bus System}
The IEEE-118 bus system is a simplified representation of the Midwest U.S. transmission system in 1962, which contains 19 generators, 35 synchronous condensers, 177 transmission lines, 9 transformers and 91 loads \cite{RChristie93a}.
Six wind farms, each with $100$ MW, are connected to bus $\{10, 25, 26, 49, 65, 66\}$, and six solar PV parks, each with an installed capacity of $60$ MW, are connected to bus $\{12, 59, 61, 80, 89, 100\}$.
Besides, there are 99 stochastic loads. Therefore, the total number of random inputs is 111. The renewable capacity penetration level is 21.96\%. 

Likewise, we first apply the power flow tool to the deterministic system. It is shown that the power flow solution is feasible. 
However, this system is prone to voltage violation since 160\% of load powers at bus $\{19, 20, 21, 43, 44, 45, 50, 51, 52\}$ are assumed, i.e., a heavy loading condition is considered to investigate the impacts of randomness on the feasibility of power flow solutions. 

\begin{figure}[h]
\centering
\includegraphics[width=0.45\textwidth,keepaspectratio=true,angle=0]{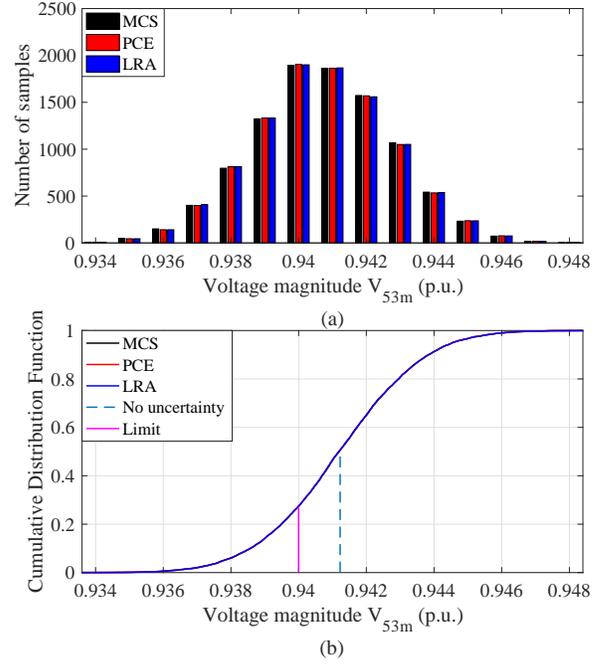}
\caption{The probability distributions of the voltage magnitude at Bus 53 estimated by the MCS, the PCE and the LRA, which are almost overlapped. Besides, there is 28\% probability that the bus voltage will be below the lower limit due to the uncertainties. (a) Probability distribution of $V_{53m}$. (b) Cumulative distribution function of $V_{53m}$.}
\label{fig:bus_v_dist}
\end{figure}

Next, we exploit the proposed LRA method to assess the probabilistic characteristics 
of the bus voltages and the branch flows 
and compare the results with those of the LHS-based MCS and with those of the sparse PCE method.
Applying the proposed algorithm, 441 simulations are needed in Step 4-5 to build up the LRAs \eqref{eq:lra_rank_r_pce} of the PPF solutions. 
Moreover, the mean and standard deviation of the responses computed in Step 6 are compared with those of the LHS-MCS and with those of the sparse PCE method as shown in Table \ref{tab:comp_mean_118} and \ref{tab:comp_var_118}. Particularly, the PDF and the CDF of the voltage magnitude at Bus 53 assessed in Step 7 are compared with those of the MCS and the sparse PCE method using 10000 samples as presented in Fig. \ref{fig:bus_v_dist}. 
All these results and comparisons clearly demonstrate that the LRA method can provide accurate estimation for the probabilistic characteristics of the PPF solutions. Particularly, the accuracy of the LRA is comparable to that of the sparse PCE method.
\begin{table}[h]
\renewcommand{\arraystretch}{1.3}
\caption{A Comparison of the estimated mean of the bus voltage and the branch flow by different methods}
\label{tab:comp_mean_118}
\centering
\begin{tabular}{c|c|c|c|c|c}
\hline
$V_{im}/S_{ij}$ & $\mu_{mcs}$ & $\mu_{pce}$ & $\mu_{lra}$ & ${\frac{\Delta \mu_{pce}}{\mu_{mcs}}} \%$ & ${\frac{\Delta \mu_{lra}}{\mu_{mcs}}} \%$ \\
%
%
\hline
$V_{53,m}$ & 0.9412 & 0.9412 & 0.9412 & -0.0002 & -0.0002 \\
\hline
$V_{21,m}$ & 0.9435 & 0.9435 & 0.9435 & 0.0008 & -0.0005 \\
\hline
$V_{44,m}$ & 0.9530 & 0.9530 & 0.9530 & 0.0049 & 0.0029 \\
\hline
$V_{20,m}$ & 0.9465 & 0.9465 & 0.9465 & 0.0004 & -0.0003 \\
\hline
$S_{49-69}$ & 0.6652 & 0.6631 & 0.6676 & -0.3243 & 0.3595 \\
\hline
$S_{47-69}$ & 0.7716 & 0.7684 & 0.7729 & -0.4229 & 0.1614 \\
\hline
\end{tabular}
\end{table}
\begin{table}[h]
\renewcommand{\arraystretch}{1.3}
\caption{A Comparison of the estimated standard deviation of the bus voltage and the branch flow by different methods }
\label{tab:comp_var_118}
\centering
\begin{tabular}{c|c|c|c|c|c}
\hline
${V_{im}/S_{ij}}$ & ${\sigma_{mcs}}$ & ${\sigma_{pce}}$ & ${\sigma_{lra}}$ & ${\frac{\Delta \sigma_{pce}}{\sigma_{mcs}} \%}$ & ${\frac{\Delta \sigma_{lra}}{\sigma_{mcs}} \%}$ \\
%
%
\hline
$V_{53,m}$ & 0.0021 & 0.0021 & 0.0021 & -0.1653 & -0.0370 \\
\hline
$V_{21,m}$ & 0.0036 & 0.0036 & 0.0036 & -1.6728 & 0.2189 \\
\hline
$V_{44,m}$ & 0.0067 & 0.0066 & 0.0068 & -2.3151 & 0.2594 \\
\hline
$V_{20,m}$ & 0.0025 & 0.0025 & 0.0025 & -1.6444 & 0.1961 \\
\hline
$S_{49-69}$ & 0.2694 & 0.2619 & 0.2697 & -2.8034 & 0.1002 \\
\hline
$S_{47-69}$ & 0.2718 & 0.2565 & 0.2719 & -5.6293 & 0.0091 \\
\hline
\end{tabular}
\end{table}

In terms of efficiency, the LHS-based MCS needs to run 10000 simulations (i.e. solving (\ref{eq:ppf_equation})), while the LRA only requires 441 simulations for solving the coefficients and the weighting factors of (\ref{eq:lra_rank_r_pce}). 
Table \ref{tab:time_compare} lists a break-down time consumption evaluating the six responses listed in Table \ref{tab:comp_mean_118}:
the time for the experimental design ${t_{ed}}$, for solving the coefficients and the weighting factors ${t_{sc}}$, for evaluating the statistic samples ${t_{es}}$, and the total time ${t_{total}}$.
The LRA method is approximately 8 times faster than the LHS-based MCS, and 1.5 times faster than the sparse PCE method. It is worth noting that the ${t_{sc}}$ and ${t_{es}}$ for the PCE and the LRA will increase linearly with the number of desired responses.
\begin{table}[h]
\renewcommand{\arraystretch}{1.3}
\caption{A Comparison of the Computation Time Required by the LHS-based MCS, the PCE and the LRA}
\label{tab:time_compare}
\centering
\begin{tabular}{c|c|c|c|c}
\hline
Method & ${t_{ed}(s)}$ & ${t_{sc}(s)}$ & ${t_{es}(s)}$ & ${t_{total}(s)}$ \\
\hline
MCS & -- & -- & 653.94 & 653.94 \\
\hline
PCE & 28.84 & 73.96 & 0.29 & 103.09 \\
\hline
LRA & 28.84 & 46.53 & 1.39 & 76.76 \\
\hline
\end{tabular}
\end{table}

Similar to the previous example, it can be observed that the uncertainty may affect the dispatching/planning scheme of the system. 
For instance, $|V_{53}|$ without considering the uncertainty is 0.9412 p.u. which is within the feasible voltage range $[0.94, 1.06]$. However, the probability corresponding to 0.94 p.u. at the CDF curve (Fig. \ref{fig:bus_v_dist}) is 0.28, indicating that there is 28\% probability that the voltage at Bus 53 will be below the lower limit. 
Such high probability of voltage violation implies that a feasible dispatch/planning scheme for the deterministic system may become unfeasible when the randomness of RES and loads are incorporated. 
\subsection{The Modified European 1354-Bus System} \label{European_1354}

The 1354-bus system is a part of the European high voltage transmission network which contains 1354 buses, 260 generators, and 1991 branches. The total load of this network is 73,060 MW and 13,401 Mvar \cite{CJosz16a}.
To assess the impact of uncertainties on the feasible power flow solution of this network, 20 wind generators, each with 100MW, and 20 solar PVs, each with 80MW, are connected to the system. All parameters of the test system are available online: \url{https://github.com/shenghao/LRA-PPF}.
There are 713 random inputs in total including the random loads.

\begin{figure}[h]
\centering
\includegraphics[width=0.45\textwidth,keepaspectratio=true,angle=0]{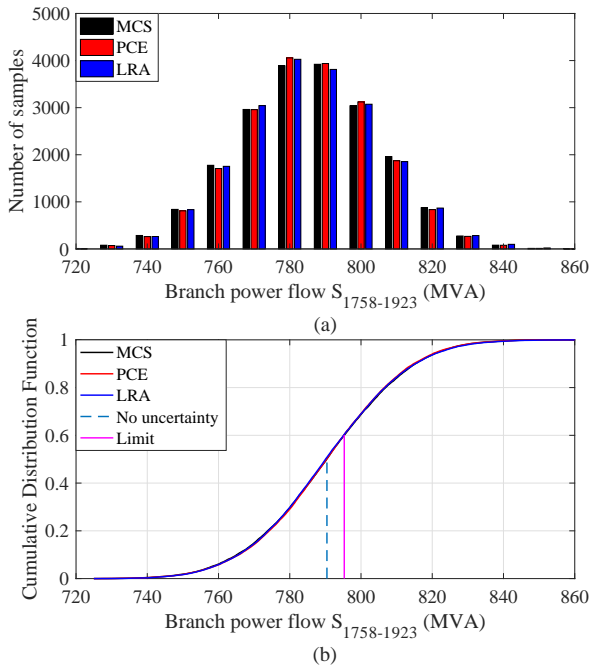}
\caption{The probability distributions of the apparent power at branch 1758-1923 estimated by the MCS, the PCE and the LRA, which are almost overlapped. Besides, there is 40\% probability that the branch will be overloaded due to the uncertainties. (a) Probability distribution of $S_{1758-1923}$. (b) Cumulative distribution function of $S_{1758-1923}$.}
\label{fig:adc_pdf_cdf_1354}
\end{figure}

This system is prone to thermal violations, hence we study the probabilistic characteristics of the line flows of the set of branches that are most overloaded. 
The estimated mean and standard deviation of the responses computed by the proposed LRA are compared with those of the LHS-based MCS as shown in Table \ref{tab:comp_mean_std_1354}.
It can be seen that the LRA method can provide accurate estimations for the statistics of the PPF solutions.
\begin{table}[h]
\renewcommand{\arraystretch}{1.3}
\caption{A Comparison of the estimated mean and standard deviation of the branch flow by different methods}
\label{tab:comp_mean_std_1354}
\centering
\resizebox{0.49\textwidth}{!}{
\begin{tabular}{>{\centering\arraybackslash}m{1.3cm}|c|c|>{\centering\arraybackslash}m{0.9cm}|c|c|>{\centering\arraybackslash}m{0.9cm}}
\hline
${V_{im}/S_{ij}}$ & ${\mu_{mcs}}$ & ${\mu_{lra}}$ & ${\frac{\Delta \mu_{lra}}{\mu_{mcs}}\%}$ & ${\sigma_{mcs}}$ & ${\sigma_{lra}}$ & ${\frac{\Delta \sigma_{lra}}{\sigma_{mcs}}\%}$ \\
%
%
\hline
$S_{118-4598}$ & 8.6613 & 8.6616 & 0.0032 & 0.6782 & 0.6779 & -0.0416 \\
\hline
$S_{1754-960}$ & 7.6526 & 7.6524 & -0.0030 & 0.5192 & 0.5197 & 0.0924 \\
\hline
$S_{6901-4874}$ & 5.6797 & 5.6797 & -0.0004 & 0.1744 & 0.1745 & 0.0891 \\
\hline
$S_{7267-6581}$ & 7.1229 & 7.1227 & -0.0019 & 0.1766 & 0.1760 & -0.3457 \\
\hline
$S_{1001-3580}$ & 5.7043 & 5.7041 & -0.0027 & 0.1596 & 0.1599 & 0.1500 \\
\hline
$S_{1001-516}$ & 5.7132 & 5.7131 & -0.0011 & 0.1179 & 0.1184 & 0.3920 \\
\hline
$S_{1758-1923}$ & 7.9033 & 7.9034 & 0.0004 & 0.1939 & 0.1934 & -0.2739 \\
\hline
\end{tabular}}
\end{table}
Similar results can be found for the estimation of the PDF and the CDF of the PPF solutions as presented in Fig. \ref{fig:adc_pdf_cdf_1354}. The proposed LRA method can provide accurate PDF and CDF approximation for the line flow at branch 1758-1923. Likewise, the deterministic solution renders 40\% probability of violating the thermal limit in the stochastic system due to the uncertainties, which evidently demonstrate the necessity of applying a probabilistic framework for power flow study. 

In power system planning, the 10\% and 90\% confidence levels that the line flow will not exceed are essential to planning engineers, since the corresponding line flow values roughly reflect the desired capacity of the transmission path \cite{PZhang04a}. As an illustration, the MVA values at 10\% (e.g., $S_{mcs}^{10}$) and 90\% (e.g., $S_{mcs}^{90}$) confidence levels of the lines in Table \ref{tab:comp_mean_std_1354} are computed using the MCS and the LRA. These results have shown that the proposed LRA method can provide planning engineers with an accurate estimate of the two indices, as shown in Table \ref{tab:conf_levels_1354}.
\begin{table}[h]
\renewcommand{\arraystretch}{1.3}
\caption{A Comparison of the estimated line flow values at the 10\% and 90\% confidence levels by different methods}
\label{tab:conf_levels_1354}
\centering
\resizebox{0.49\textwidth}{!}{
\begin{tabular}{c|c|c|c|c|c|c}
\hline
${S_{ij}}$ & ${S_{mcs}^{10}}$ & ${S_{lra}^{10}}$ & ${\frac{\Delta S_{lra}^{10}}{S_{mcs}^{10}}\%}$ & ${S_{mcs}^{90}}$ & ${S_{lra}^{90}}$ & ${\frac{\Delta S_{lra}^{90}}{S_{mcs}^{90}}\%}$ \\
%
%
\hline
$S_{118-4598}$ & 7.8045 & 7.8116 & 0.0911 & 9.5266 & 9.5349 & 0.0873 \\
\hline
$S_{1754-960}$ & 6.9978 & 7.0051 & 0.1042 & 8.3152 & 8.3198 & 0.0561 \\
\hline
$S_{6901-4874}$ & 5.4569 & 5.4584 & 0.0282 & 5.9048 & 5.9057 & 0.0139 \\
\hline
$S_{7267-6581}$ & 6.8885 & 6.8941 & 0.0809 & 7.3523 & 7.3501 & -0.0306 \\
\hline
$S_{1001-3580}$ & 5.4841 & 5.4957 & 0.2099 & 5.8998 & 5.9061 & 0.1067 \\
\hline
$S_{1001-516}$ & 5.5529 & 5.5564 & 0.0633 & 5.8587 & 5.8627 & 0.0683 \\
\hline
$S_{1758-1923}$ & 7.6511 & 7.6560 & 0.0638 & 8.1527 & 8.1526 & -0.0009 \\
\hline
\end{tabular}}
\end{table}

In terms of simulation effort, the LHS-base MCS requires 50000 simulations to obtain a converged result, while the proposed LRA method requires only 3566 simulations to get comparable accuracy. 
It should be noted that the sparse PCE is unable to solve this problem (with 713 inputs) due to the large size of the expanded multivariate polynomials. The issue of `out of memory' is met on the computer (DELL OptiPlex 7050 with Intel Core i7-7700 (3.6GHz), 16GB RAM).

To make a more detailed comparison between the sparse PCE and the LRA regarding the capability of handling high-dimensional problems, six reduced systems with an increasing number of random inputs (i.e., 100, 200, 300, 400, 500, 600) are created based on the 1354-bus system by fixing small loads to their mean values.
Again, we apply the MCS, the sparse PCE and the LRA to evaluate their PPF solutions. 

Table \ref{tab:number_of_unknowns} shows the number of unknowns to be solved in the sparse PCE and the LRA. Even with the sophisticated hyperbolic truncation scheme \cite{GBlatman09a}, the number of unknown coefficients to be sent into the least angle regression algorithm \cite{BEfron04a} (column 4, $N_{trunc}^{pce}$) grows dramatically as the number of inputs increases. Nevertheless, the unknown coefficients plus weighting factors in LRA (column 8, $N_{solved}^{lra}$) grows only linearly. When the number of inputs reach 600, the spare PCE failed due to `out of memory'. These results demonstrate the advantage of the LRA method over the PCE method on handling high-dimensional problems when the number of random inputs is enormous.
\begin{table}[h]
\renewcommand{\arraystretch}{1.3}
\caption{Comparison of the unknowns to be solved by the PCE and the LRA for different number of inputs}
\label{tab:number_of_unknowns}
\centering
\resizebox{0.49\textwidth}{!}{
\begin{tabular}{c|>{\centering\arraybackslash}m{0.5cm}|>{\centering\arraybackslash}m{1.3cm}|>{\centering\arraybackslash}m{0.8cm}|>{\centering\arraybackslash}m{0.9cm}|>{\centering\arraybackslash}m{0.5cm}|>{\centering\arraybackslash}m{0.4cm}|>{\centering\arraybackslash}m{0.8cm}}
\hline
\multirow{2}{*}{$n$} & \multicolumn{4}{c|}{Sparse PCE} & \multicolumn{3}{c}{LRA} \\
\cline{2-8}
 & $p_{max}^{pce}$ & $N_{full}^{pce}$ & $N_{trunc}^{pce}$ & $N_{solved}^{pce}$ & $p_{max}^{lra}$ & $r^{lra}$ & $N_{solved}^{lra}$ \\
\hline
100 & 3 & 176851 & \textbf{5251} & 89 & 2 & 1 & \textbf{301} \\
\hline
200 & 3 & 1373701 & \textbf{20501} & 161 & 2 & 1 & \textbf{601} \\
\hline
300 & 4 & 348881876 & \textbf{135751} & 200 & 2 & 1 & \textbf{901} \\
\hline
400 & 3 & 10827401 & \textbf{81001} & 214 & 2 & 1 & \textbf{1201} \\
\hline
500 & 4 & 2656615626 & \textbf{376251} & 339 & 2 & 1 & \textbf{1501} \\
\hline
600 & - & - & - & - & 2 & 1 & \textbf{1801} \\
\hline
713 & - & - & - & - & 2 & 1 & \textbf{2140} \\
\hline
\end{tabular}}
\begin{tablenotes}
\item * $n$ is the number of inputs; $p_{max}^{pce}$ is the highest degree of multivariate polynomials;
$N_{full}^{pce}$, $N_{trunc}^{pce}$ and $N_{solved}^{pce}$ are the number of coefficients (multivariate polynomial terms) in the full PCE, the truncated PCE and the solved sparse PCE, respectively; $r^{lra}$, $p_{max}^{lra}$ and $N_{solved}^{lra}$ are the respective rank, degree and total number of coefficients plus weighting factors in the solved LRA.
The main computational effort of the PCE and the LRA are determined by $N_{trunc}^{pce}$ and $N_{solved}^{lra}$ respectively.
\item * The dash "-" indicates that the PCE encounters 'out of memory'.
\end{tablenotes}
\end{table}
Table \ref{tab:comp_capab_mean_1354} and \ref{tab:comp_capab_std_1354} show the mean and standard deviation of the line flow at the branch 1758-1923 in the six reduced systems as well as the original system.
Clearly, the LRA is able to provided accurate estimation for all of the seven systems.
\begin{table}[h]
\renewcommand{\arraystretch}{1.3}
\caption{A Comparison of the estimated mean of the line flow of branch 1758-1923 in different cases}
\label{tab:comp_capab_mean_1354}
\centering
\begin{tabular}{c|c|c|c|c|c}
\hline
$n$ & ${\mu_{mcs}}$ & ${\mu_{pce}}$ & ${\mu_{lra}}$ & ${\frac{\Delta \mu_{pce}}{\mu_{mcs}}\%}$ & ${\frac{\Delta \mu_{lra}}{\mu_{mcs}}\%}$ \\
%
%
\hline
100 & 7.9026 & 7.9030 & 7.9027 & 0.0044 & 0.0008 \\
\hline
200 & 7.9027 & 7.9022 & 7.9027 & -0.0059 & -0.0002 \\
\hline
300 & 7.9029 & 7.9029 & 7.9030 & 0.0006 & 0.0007 \\
\hline
400 & 7.9030 & 7.9030 & 7.9031 & 0.0001 & 0.0014 \\
\hline
500 & 7.9032 & 7.9032 & 7.9033 & -0.0005 & 0.0011 \\
\hline
600 & 7.9034 & - & 7.9033 & - & -0.0002 \\
\hline
713 & 7.9033 & - & 7.9034 & - & 0.0004 \\
\hline
\end{tabular}
\end{table}
\begin{table}[h]
\renewcommand{\arraystretch}{1.3}
\caption{A Comparison of the estimated standard deviation of the line flow of branch 1758-1923 in different cases}
\label{tab:comp_capab_std_1354}
\centering
\begin{tabular}{c|c|c|c|c|c}
\hline
$n$ & ${\sigma_{mcs}}$ & ${\sigma_{pce}}$ & ${\sigma_{lra}}$ & ${\frac{\Delta \sigma_{pce}}{\sigma_{mcs}} \%}$ & ${\frac{\Delta \sigma_{lra}}{\sigma_{mcs}} \%}$ \\
%
%
\hline
100 & 0.1546 & 0.1501 & 0.1547 & -2.9228 & 0.0594 \\
\hline
200 & 0.1793 & 0.1731 & 0.1795 & -3.4811 & 0.1250 \\
\hline
300 & 0.1892 & 0.1824 & 0.1890 & -3.5640 & -0.1099 \\
\hline
400 & 0.1952 & 0.1900 & 0.1941 & -2.6852 & -0.5928 \\
\hline
500 & 0.1948 & 0.1897 & 0.1947 & -2.6008 & -0.0623 \\
\hline
600 & 0.1961 & - & 0.1962 & - & 0.0447 \\
\hline
713 & 0.1939 & - & 0.1934 & - & -0.2739 \\
\hline
\end{tabular}
\end{table}

\noindent Remark: 
Since independent LRA has to be built for each response of interests, the computational effort will increase linearly with the number of desired responses.

%
\section{Conclusion and Perspectives}
In this paper, we have developed a novel method to accurately and efficiently assess PPF. Particularly, the proposed method can build up a statistically-equivalent surrogate model (i.e., the low-rank approximations) for the PPF solutions through a small number of power flow evaluations. 
Numerical studies show that
\begin{itemize}
\item The proposed LRA method can accurately estimate the probabilistic characteristics of the bus voltages and line flows in the PPF problem. 
\item The proposed LRA method is more computationally efficient 
compared to the LHS-based MCS. Moreover, once the LRA \eqref{eq:lra_rank_r_pce} for the desired response is built up, the response of any new samples can be evaluated efficiently by directly substituting to \eqref{eq:lra_rank_r_pce} instead of solving the original PPF equations \eqref{eq:ppf_equation}.
\item The proposed LRA method has higher capability of dealing high-dimensional problems compared to the PCE method. 
This merit stems from exploiting the retained tensor-product form and employing a sequential correction-updating scheme.
\end{itemize}

It has been revealed in our study that a feasible power flow may become unfeasible with a relatively high probability when the uncertainties are incorporated. The probability may be even larger if the penetration of RES increases thus need to be carefully investigated. 
In the future, we plan to develop a grouping scheme to accelerate the LRA calculation by reducing the iterations required in the correction step.

%
%

%
%



\begin{IEEEbiography}[{\includegraphics[width=1in,height=1.25in,clip,keepaspectratio]{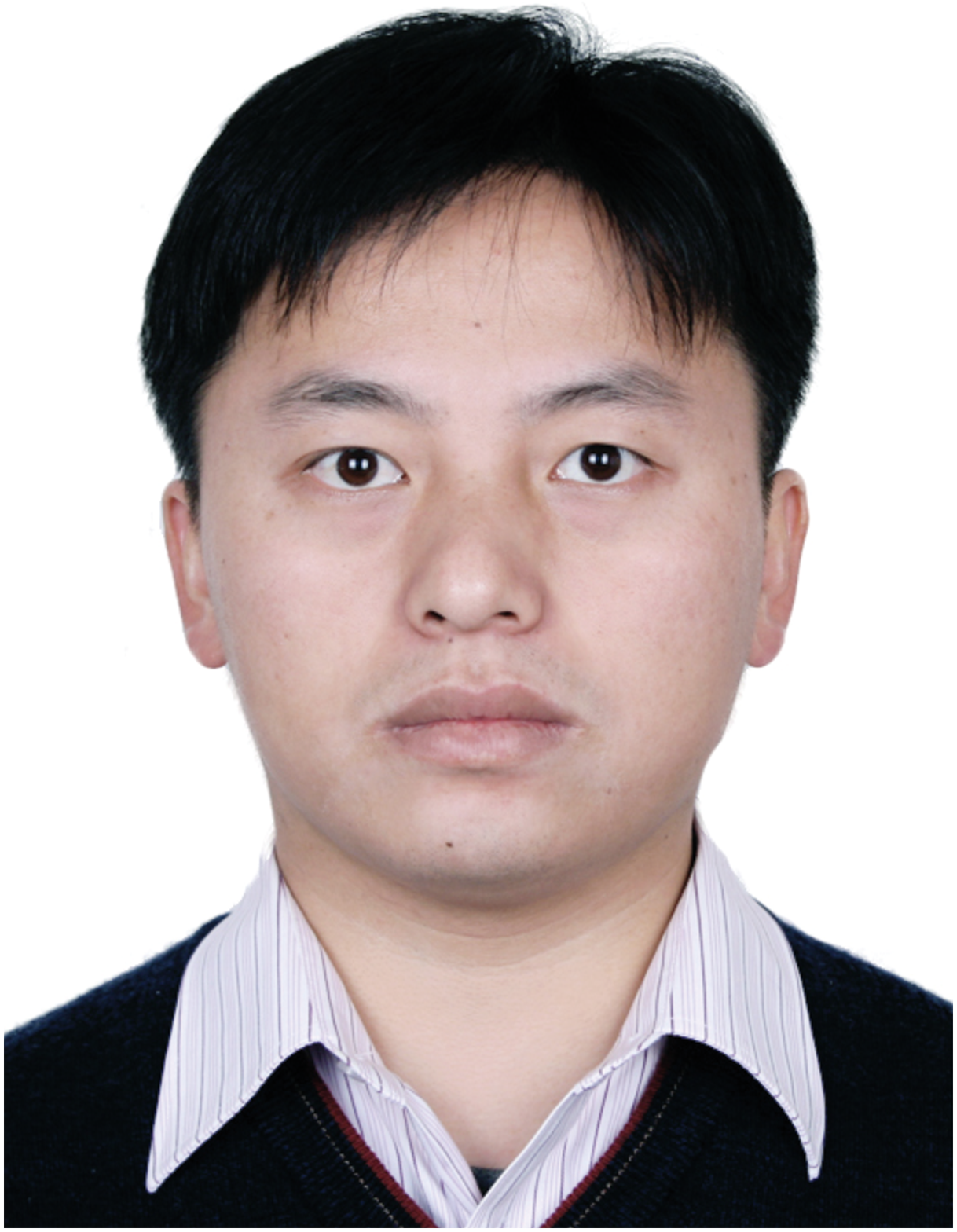}}]{Hao Sheng}
(M'14) is currently an postdoctoral fellow in the Department of Electrical and Computer Engineering at McGill University, Montreal, QC, Canada. He received the Ph.D. degree in the School of Electrical and Automation Engineering from Tianjin University, Tianjin, China, in 2014, the M.S. degree from Northeast Electric Power University, Jilin, China, in 2007 and the B.E. degree from North China Electric Power University, Baoding, China, in 2003, all in Electrical Engineering. From 2007 to 2012, he was affiliated with R\&D Centre of Beijing SiFang Automation Co., Ltd., Beijing, China, working on the development of PMU data-enhanced applications for Energy Management System (EMS) and Dynamic Security Assessment (DSA). From 2014 to 2017, he was a postdoctoral fellow in the School of Electrical and Computer Engineering at Cornell University, Ithaca, NY, USA. His research interests are in power system stability analysis and simulation, uncertainty quantification and control and their applications in power system static and dynamic security assessment.
\end{IEEEbiography}

\begin{IEEEbiography}[{\includegraphics[width=1in,height=1.25in,clip,keepaspectratio]{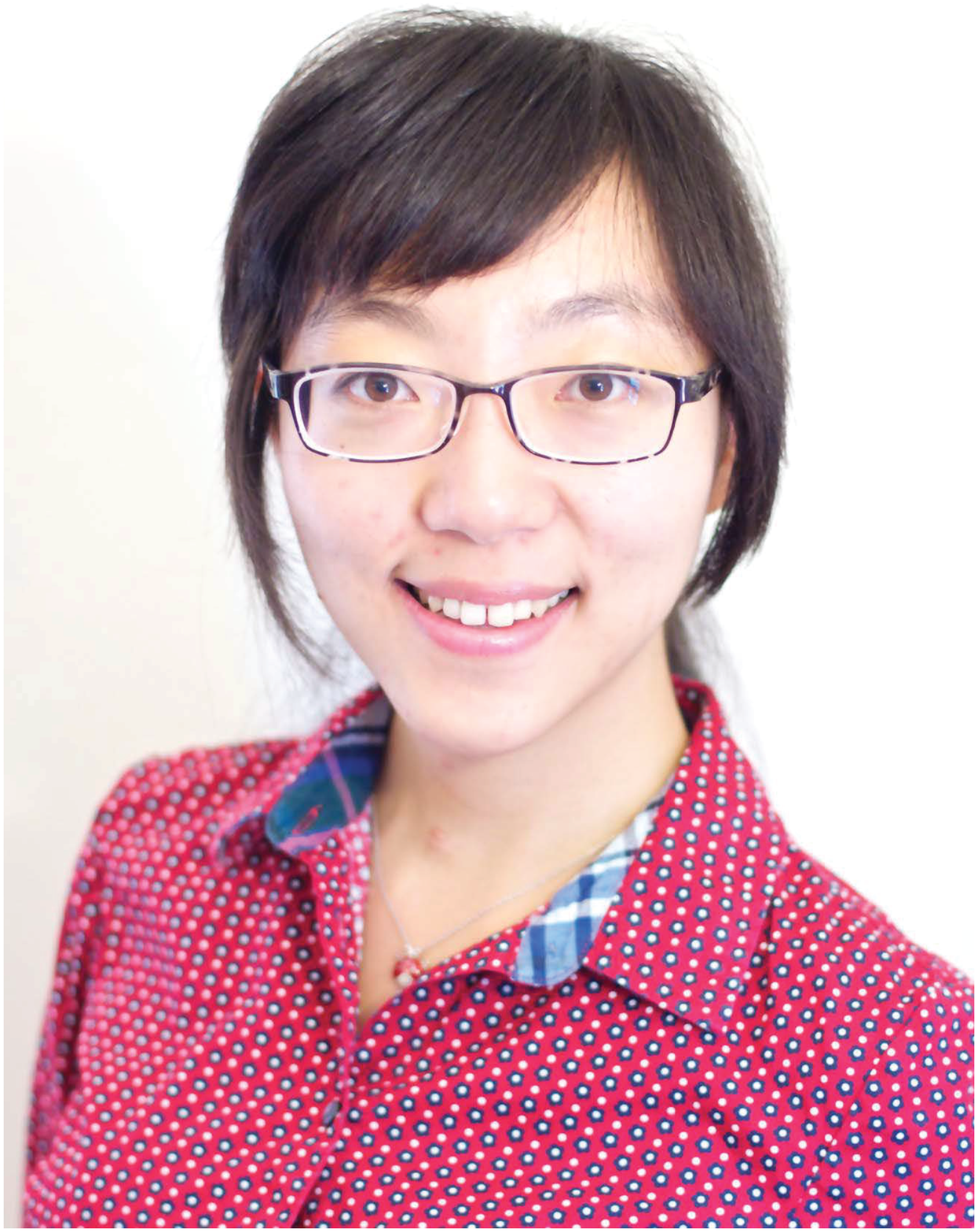}}]{Xiaozhe Wang}
is currently an Assistant Professor in the Department of Electrical and Computer Engineering at McGill University, Montreal, QC, Canada. She received the Ph.D. degree in the School of Electrical and Computer Engineering from Cornell University, Ithaca, NY, USA, in 2015, and the B.S. degree in Information Science \& Electronic Engineering from Zhejiang University, Zhejiang, China, in 2010. Her research interests are in the general areas of power system stability and control, uncertainty quantification in power system security and stability, and wide-area measurement system (WAMS)-based detection, estimation, and control.
\end{IEEEbiography}


\begin{thebibliography} {1}

\bibitem{BBorkowska74a}
B. Borkowska, "Probabilistic load flow," \textit{IEEE Trans. Power Appar. Syst.}, vol. 93, no. 3, pp. 752--759, 1974.
%
\bibitem{RNAllan74a}
R. N. Allan, B. Borkowska, and C. H. Grigg, "Probabilistic analysis of power flows," \textit{Proc. Inst. Electr. Eng.}, vol. 121, no. 12, pp. 1551--1556, 1974.
%
\bibitem{HYu09a}
H. Yu, C. Y. Chung, K. P. Wong, and J. H. Zhang, "Probabilistic load flow evaluation with hybrid Latin Hypercube sampling and Cholesky decomposition," \textit{IEEE Trans. on Power Syst.}, vol. 24, no. 2, pp. 661--667, 2009.
%
\bibitem{JHuang11a}
J. Huang, Y. Xue, Z. Y. Dong, and K. P. Wong, "An adaptive importance sampling method for probabilistic optimal power flow," in \textit{2011 IEEE Power and Energy Society General Meeting}, Detroit, MI, USA, 24--29 July 2011, pp. 1--6.
%
\bibitem{MHajian13a}
M. Hajian, W. D. Rosehart, and H. Zareipour, "Probabilistic power flow by Monte Carlo simulation with Latin Supercube sampling," \textit{IEEE Trans. Power Syst.}, vol. 28, no. 2, pp. 1550--1559, 2013.
%
\bibitem{PZhang04a}
P. Zhang and S. T. Lee, "Probabilistic load flow computation using the method of combined cumulants and Gram-Charlier expansion," \textit{IEEE Trans. on Power Syst.}, vol. 19, no. 1, pp. 676--682, 2004.
%
\bibitem{CLSu05a}
C. L. Su, "Probabilistic load-flow computation using point estimate method," \textit{IEEE Trans. on Power Syst.}, vol. 20, no. 4, pp. 1843--1851, 2005.
%
\bibitem{JMMorales07a}
J. M. Morales and J. Perez-Ruiz, "Point estimate schemes to solve the probabilistic power flow," \textit{IEEE Trans. Power Syst.}, vol. 22, no. 4, pp. 1594--1601, 2007.
%
\bibitem{FJRuizRodriguez12a}
F. J. Ruiz-Rodriguez, J. C. Hernandez, and F. Jurado, "Probabilistic load flow for photovoltaic distributed generation using the Cornish-Fisher expansion," \textit{Electr. Power Syst. Res.}, vol. 89, pp. 129--138, 2012.
%
\bibitem{MFan12a}
M. Fan, V. Vittal, G. T. Heydt, and R. Ayyanar, "Probabilistic power flow studies for transmission systems with photovoltaic generation using cumulants," \textit{IEEE Trans. Power Syst.}, vol. 27, no. 4, pp. 2251--2261, 2012
%
\bibitem{ZYRen16a}
Z. Ren, W. Li, R. Billinton, and W. Yan, "Probabilistic power flow analysis based on the stochastic response surface method," \textit{IEEE Trans. Power Syst.}, vol. 31, no. 3, pp. 2307--2315, 2016.
%
\bibitem{FNi17a}
F. Ni, P. H. Nguyen, and J. F. G. Cobben, "Basis-adaptive sparse polynomial chaos expansion for probabilistic power flow," \textit{IEEE Trans. Power Syst.}, vol. 32, no. 1, pp. 694--705, 2017.
%
\bibitem{EHaesen09a}
E. Haesen, C. Bastiaensen, J. Driesen, and R. Belmans, "A probabilistic formulation of load margins in power systems with stochastic generation," \textit{IEEE Trans. Power Syst.}, vol. 24, no. 2, pp. 951--958, 2009.
%
\bibitem{HSheng18a}
H. Sheng and X. Wang, "Applying polynomial chaos expansion to assess probabilistic available delivery capability for distribution networks with renewables," \textit{IEEE Trans. Power Syst.}, 2018, (Early Access).
%
\bibitem{KKonakli16a}
K. Konakli and B. Sudret, "Polynomial meta-models with canonical low-rank approximations: numerical insights and comparison to sparse polynomial chaos expansions," \textit{J. Comput. Phys.}, vol. 321, pp. 1144--1169, 2016.
%
\bibitem{FHitchcock27a}
F. Hitchcock, "The expression of a tensor or a polyadic as a sum of products," \textit{J. Math. Phys.}, vol. 6, pp. 164--189, 1927.
%
\bibitem{ADoostan13a}
A. Doostan, A. Validi, and G. Iaccarino, "Non-intrusive low-rank separated approximation of high-dimensional stochastic models," \textit{Comput. Methods Appl. Mech. Eng.}, vol. 263, pp. 42--55, 2013.
%
\bibitem{MChevreuil15a}
M. Chevreuil, R. Lebrun, A. Nouy, and P. Rai, "A least-squares method for sparse low-rank approximation of multivariate functions," \textit{SIAM/ASA J. Uncertain. Quantificat.}, vol. 3, no. 1, pp. 897--921, 2015.
%
\bibitem{JCarpentier62a}
J. Carpentier, "Contribution a letude du dispatching economique," \textit{Bull. Soc. Francaise Electricians}, vol. 8, pp. 431--447, 1962.
%
\bibitem{HWDommel68a}
H. W. Dommel and W. F. Tinney, "Optimal power flow solutions," \textit{IEEE Trans. Power App. Syst.}, vol. PAS-87, no. 10, pp. 1866--1876, 1968.
%
\bibitem{MMadrigal98a}
M. Madrigal, K. Ponnambalam, and V. H. Quintana, "Probabilistic optimal power flow," in \textit{IEEE Canadian Conf. Electrical Computer Engineering}, Waterloo, ON, Canada, 1998, pp. 385--388.
%
\bibitem{HZhang11a}
H. Zhang and P. Li, "Chance constrained programming for optimal power flow under uncertainty," \textit{IEEE Trans. Power Syst.}, vol. 26, no. 4, pp. 2417--2424, 2011.
%
\bibitem{SBahrami18a}
S. Bahrami, M. H. Amini, M. Shafie-Khah, and J. P. S. Catalao, "A decentralized renewable generation management and demand response in power distribution networks," \textit{IEEE Trans. Sustain. Energy}, vol. 9, no. 0, pp. 1783--1797, 2018.
%
\bibitem{SHKaraki99a}
S. H. Karaki, R. B. Chedid, and R. Ramadan, "Probabilistic performance assessment of autonomous solar-wind energy conversion systems," \textit{IEEE Trans. Energy Conver.}, vol. 14, no. 3, pp. 766--772, 1999.
%
\bibitem{Xiaozhe15}
X. Z. Wang, H. D. Chiang, J. H. Wang, H. Liu, and T. Wang, "Long-term stability analysis of power systems with wind power based on stochastic differential equations: model development and foundations," \textit{IEEE Trans. on Sustain. Energy}, vol. 6, no. 4, pp. 1534--1542, 2015.
%
\bibitem{Xiaozhe17}
X. Z. Wang, T. Wang, H. D. Chiang, J. H. Wang, and H. Liu, "A framework for dynamic stability analysis of power systems with volatile wind power," in \textit{IEEE Journal on Emerging and Selected Topics in Circuits and Systems}, 2017.
%
\bibitem{SHJangamshetti99a}
S. H. Jangamshetti and V. G. Rau, "Site matching of wind turbine generators: a case study," \textit{IEEE Trans. Energy Conver.}, vol. 14, no. 4, pp. 1537--1543, 1999.
%
\bibitem{SAAkdag09a}
S. A. Akdag and A. Dinler, "A new method to estimate Weibull parameters for wind energy applications," \textit{Energy Conversion and Management}, vol. 50, pp. 1761--1766, 2009.
%
\bibitem{XRan16a}
X. Ran and S. Miao, "Three-phase probabilistic load flow for power system with correlated wind, photovoltaic and load," \textit{IET Gener., Transm. \& Distrib.}, vol. 10, no. 12, pp. 3093--3101, 2016.
%
\bibitem{SRoy02a}
S. Roy, "Market constrained optimal planning for wind energy conversion systems over multiple installation sites," \textit{IEEE Trans. Energy Convers.}, vol. 17, no. 1, pp. 124--129, 2002.
%
\bibitem{EHCamm09b}
E. H. Camm, M. R. Behnke, O. Bolado, M. Bollen, M. Bradt, C. Brooks, W. Dilling, M. Edds, W. J. Hejdak, D. Houseman, S. Klein, F. Li, J. Li, P. Maibach, T. Nicolai, J. Patino, S. V. Pasupulati, N. Samaan, S. Saylors, T. Siebert, T. Smith, M. Starke, , and R. Walling, "Characteristics of wind turbine generators for wind power plants," in \textit{IEEE Power \& Energy Society General Meeting}, Calgary, AB, Canada, 26-30 July 2009, pp. 1--5.
%
\bibitem{FYEttoumi02a}
F. Y. Ettoumi, A. Mefti, A. Adane, and M. Y. Bouroubi, "Statistical analysis of solar measurements in Algeria using beta distributions," \textit{Renewable Energy}, vol. 26, no. 1, pp. 47--67, 2002.
%
\bibitem{CBOwen08a}
C. B. Owen, "Parameter estimation for the beta distribution," Master's thesis, Department of Statistics, Provo, UT, USA, 2008.
%
\bibitem{WECC10}
WECC. WECC guide for representation of photovoltaic systems in large-scale load flow simulations. [Online]. Available: https://www.wecc.biz
%
\bibitem{AEllis12a}
A. Ellis, R. Nelson, E. V. Engeln, R. Walling, J. MacDowell, L. Casey, E. Seymour, W. Peter, C. Barker, B. Kirby, and J. R. Williams, "Reactive power performance requirements for wind and solar plants," in \textit{2012 IEEE Power and Energy Society General Meeting}, San Diego, CA, USA,22--26 July 2012, pp. 1--8.
%
\bibitem{RBillinton08a}
R. Billinton and D. Huang, "Effects of load forecast uncertainty on bulk electric system reliability evaluation," \textit{IEEE Trans. Power Syst.}, vol. 23, no. 2, pp. 418--425, 2008.
%
\bibitem{WYLi13a}
W. Li, E. Vaahedi, and Z. Lin, "BC Hydro's transmission reliability margin assessment in total transfer capability calculations," \textit{IEEE Trans. Power Syst.}, vol. 28, no. 4, pp. 4796--4802, 2013.
%
\bibitem{DBXiu02a}
D. B. Xiu and G. E. Karniadakis, "The Wiener--Askey polynomial chaos for stochastic differential equations," \textit{SIAM J. Sci. Comput.}, vol. 24, no. 2, pp. 619--644, 2002.
%
\bibitem{RLebrun09b}
R. Lebrun and A. Dutfoy, "A generalization of the Nataf transformation to distributions with elliptical copula," \textit{Prob. Eng. Mech.}, vol. 24, no. 2, pp. 172--178, 2009.
%
\bibitem{SMarelli17a}
S. Marelli and B. Sudret, "UQLab user manual--polynomial chaos expansions," Chair of Risk, Safety \& Uncertainty Quantification, ETH Zurich, Tech. Rep., 2017, UQLab-V1.0-104.
%
\bibitem{ZMSalameh95a}
Z. M. Salameh, B. S. Borowy, and A. R. A. Amin, "Photovoltaic module-site matching based on the capacity factors," \textit{IEEE Trans. Energy Convers.}, vol. 10, no. 2, pp. 326--332, 1995.
%
\bibitem{SOladyshkin12a}
S. Oladyshkin and W. Nowak, "Data-driven uncertainty quantification using the arbitrary polynomial chaos expansion," \textit{Reliability Engineering and System Safety}, vol. 106, pp. 179--190, 2012.
%
\bibitem{PRai14a}
P. Rai, "Sparse low-rank approximation of multivariate functions–applications in uncertainty quantification," Ph.D. dissertation, Engineering Sciences [Physics], Ecole Centrale Nantes, 2014.
%
\bibitem{RLebrun09a}
R. Lebrun and A. Dutfoy, "An innovating analysis of the Nataf transformation from the copula viewpoint," \textit{Prob. Eng. Mech.}, vol. 24, no. 3, pp. 312--320, 2009.
%
\bibitem{MATPOWER}
R. D. Zimmerman, C. E. Murillo-Sanchez, and R. J. Thomas, "MATPOWER: steady-state operations, planning and analysis tools for power systems research and education," \textit{IEEE Trans. Power Syst.}, vol. 26, no. 1, pp. 12--19, 2011.
%
\bibitem{SMarelli14a}
S. Marelli and B. Sudret, "UQLab: A framework for uncertainty quantification in matlab," in \textit{Proc. 2nd Int. Conf. on Vulnerability, Risk Analysis and Management (ICVRAM2014)}, Liverpool, United Kingdom, 2014, pp. 2554--2563.
%
\bibitem{RChristie93a}
R. Christie. Power systems test case archive. University of Washington, Seattle, WA, USA. [Online]. Available: https://www2.ee.washington. edu/research/pstca/pf118/pg\ tca118bus.htm
%
\bibitem{CJosz16a}
C. Josz, S. Fliscounakis, J. Maeght, and P. Panciatici. Data in MATPOWER and QCQP format: iTesla, RTE snapshots, and PEGASE. Accessed on 2018-09-10. [Online]. Available: http://arxiv.org/abs/1603. 01533
%
\bibitem{GBlatman09a}
G. Blatman, "Adaptive sparse polynomial chaos expansions for uncertainty propagation and sensitivity analysis," Ph.D. dissertation, Mechanical Engineering, Clermont-Ferrand, France, 2009.
%
\bibitem{BEfron04a}
B. Efron, T. Hastie, I. Johnstone, and R. Tibshirani, "Least angle regression," \textit{The Annual of Statistics}, vol. 32, no. 2, pp. 407--499, 2004.

\end{thebibliography}
\end{document}